\documentclass[aps,prd,groupedaddress,superscriptaddress,byrevtex,a4paper]{revtex4}

\pdfoutput=1

\usepackage[utf8]{inputenc}
\usepackage[T1]{fontenc}

\usepackage{amssymb,amsfonts,amsmath,amsthm}             %AMS packages
\usepackage{slashed}

\usepackage{graphicx}                                       % Include figure files
\usepackage[usenames]{color}
\usepackage[tight,TABTOPCAP,hang,raggedright,normalsize]{subfigure}   % for more pictures in one figure

\usepackage{multirow}
\usepackage{dcolumn}

\usepackage[
            pdftitle={Nucleon distribution amplitudes from lattice QCD},
            pdfauthor={Nikolaus Warkentin},
            pdfsubject={Nucleon distribution amplidtudes and proton decay
matrix elements on the lattice},
            pdfkeywords={moments, hadron, nucleon, baryon, quark,
distribution,wave,function},
            pdfpagemode=UseOutlines,
            colorlinks=true,
            pdfproducer={University of Regensburg},
            ]{hyperref}

\newcommand{\dev}{$\chi^2/\textrm{d.o.f}$}
\newcommand{\msb}{$\overline{\text{MS}}$}
\newcommand{\imi}{i}
\DeclareRobustCommand\openone{\leavevmode\hbox{\small1\normalsize\kern-.33em1}}     % Identity Matrix Declaration

\begin{document}

\vspace*{-2.0cm} 
\begin{flushright}
DESY 08-166 \\ Edinburgh 2008/45
\end{flushright}

\title{Nucleon distribution amplitudes and proton decay matrix elements on the lattice}

\author{Vladimir M. \surname{Braun}}
    \affiliation{Institut f\"ur Theoretische Physik, Universit\"at Regensburg, 93040 Regensburg, Germany}
\author{Meinulf \surname{G\"ockeler}}
    \affiliation{Institut f\"ur Theoretische Physik, Universit\"at Regensburg, 93040 Regensburg, Germany}
\author{Roger \surname{Horsley}}
    \affiliation{School of Physics, University of Edinburgh, Edinburgh EH9~3JZ, UK}
\author{Thomas \surname{Kaltenbrunner}}
    \affiliation{Institut f\"ur Theoretische Physik, Universit\"at Regensburg, 93040 Regensburg, Germany}
\author{Yoshifumi \surname{Nakamura}}
    \affiliation{Deutsches Elektronen-Synchrotron DESY and John von Neumann Institut f\"ur
Computing NIC, 15738 Zeuthen, Germany}
\author{Dirk \surname{Pleiter}}
    \affiliation{Deutsches Elektronen-Synchrotron DESY and John von Neumann Institut f\"ur
Computing NIC, 15738 Zeuthen, Germany}
\author{\firstname{Paul} E.~L. \surname{Rakow}}
    \affiliation{Theoretical Physics Division, Department of Mathematical Sciences, University of Liverpool, Liverpool L69~3BX, UK}
\author{Andreas \surname{Sch\"afer}}
    \affiliation{Institut f\"ur Theoretische Physik, Universit\"at Regensburg, 93040 Regensburg, Germany}
    \affiliation{Yukawa Institute for Theoretical Physics, Kyoto University, Japan}
\author{Gerrit \surname{Schierholz}}
    \affiliation{Institut f\"ur Theoretische Physik, Universit\"at Regensburg, 93040 Regensburg, Germany}
    \affiliation{Deutsches Elektronen-Synchrotron DESY and John von Neumann Institut f\"ur
Computing NIC, 15738 Zeuthen, Germany}
\author{Hinnerk~St\"uben}
    \affiliation{Konrad-Zuse-Zentrum f\"ur Informationstechnik Berlin, 14195 Berlin, Germany}

\author{Nikolaus \surname{Warkentin}}
    \affiliation{Institut f\"ur Theoretische Physik, Universit\"at Regensburg, 93040 Regensburg, Germany}
\author{\firstname{James} M. \surname{Zanotti}}
    \affiliation{School of Physics, University of Edinburgh, Edinburgh EH9~3JZ, UK}
\collaboration{QCDSF Collaboration}\noaffiliation

\begin{abstract}

Baryon distribution amplitudes (DAs) are crucial for the theory of hard
exclusive reactions. We present a calculation of the first few moments
of the leading-twist nucleon DA within lattice QCD. In addition we 
deal with the normalization of the next-to-leading (twist-four) DAs.
The matrix elements determining the latter quantities are also
responsible for proton decay in Grand Unified Theories. Our lattice
evaluation makes use of gauge field configurations generated with
two flavors of clover fermions. The relevant operators are renormalized
nonperturbatively with the final results given in the \msb\ scheme.
We find that the deviation of the leading-twist nucleon DA from its 
asymptotic form is less pronounced than sometimes claimed in the
literature.

\end{abstract}

\maketitle

\section{Introduction}

The notion of baryon distribution amplitudes (DAs) refers to the valence 
component of the Bethe-Salpeter wave function at small transverse 
separations and is central for the theory of hard exclusive reactions
involving baryons~\cite{Lepage:1979za,Brodsky:1980sx,Brodsky:1981kj,
Chernyak:1977fk,Efremov:1979qk,Lepage:1979zb,Lepage:1980fj,
Chernyak:1984bm,Chernyak:1987nu}.
As usual for a field theory, extraction of the asymptotic behavior 
(in our case for vanishing transverse separation) introduces divergences
that can be studied by the renormalization-group (RG) method.
The distribution amplitude $\varphi$ thus becomes a function of the three 
quark momentum fractions $x_i$ and the scale that serves as a UV cutoff
in the allowed transverse momenta. Solving the corresponding RG 
equations in leading logarithmic accuracy~\cite{Peskin:1979mn,Brodsky:1980ny}
one is led to the expansion
\begin{equation}
 \varphi(x_i,\mu^2)=120 x_1 x_2 x_3 \sum_{n=0}^\infty \sum_{l=0}^n c_{nl}(\mu_0) P_{nl}(x_i) 
     \left(\frac{\alpha_s(\mu)}{\alpha_s(\mu_0)}\right)^{\gamma_{nl}/\beta_0}.
     \label{eq_daexpansion1}
\end{equation} 
The summation goes over all multiplicatively renormalizable 
operators built of three quarks and $n$ derivatives and $\beta_0$ is the first
coefficient of the beta function. The polynomials 
$P_{nl}(x_i)$ and anomalous dimensions $\gamma_{nl}$ are obtained by 
diagonalizing the mixing matrix for the three-quark operators
\begin{displaymath}
(D_+^{k_1} q) (D_+^{k_2} q)(D_+^{k_3} q)\; , \;
  k_1+k_2+k_3 =n \;,
\end{displaymath}
and the $c_{nl}(\mu_0)$ are the corresponding (nonperturbative) matrix 
elements.

The theory of nucleon DAs has reached a certain degree of maturity. 
In particular the scale dependence is well 
understood~\cite{Braun:1999te,Braun:2008ia}
and it reveals important symmetries 
of the quantum theory that are not seen at the level of the QCD Lagrangian
\cite{Braun:1998id}. At the same time, they are much less studied as 
compared to the usual parton distributions. 
One reason is that the approach to the perturbative factorization regime in
hard reactions appears to be slow. There is overwhelming evidence that, e.g., 
electromagnetic and transition form factors at currently available 
momentum transfers of the order of a few GeV$^2$  
\cite{Jones:1999rz,Gayou:2001qt,Gayou:2001qd,Punjabi:2005wq} receive 
large nonfactorizable contributions from large transverse distances,
usually referred to as soft (Feynman) or end-point contributions, and
possibly from higher-twist corrections. This is indicated, for
example, by the fact that the helicity selection rules are strongly 
violated. Another reason is that nucleon DAs enter physical observables 
in a rather complicated way through convolution integrals, 
integrated with smooth functions of the momentum fractions. This makes 
an experimental determination of the DAs pointwise in $x_i$ very difficult. 
A qualitative picture suggested by QCD sum rule calculations is that 
the valence quark with the spin parallel to that of the proton carries 
most of its momentum \cite{Chernyak:1984bm,King:1986wi,Chernyak:1987nu}.
It is timely to make this picture quantitative; lattice QCD ist best
suited for this purpose \cite{Martinelli:1988xs,Gavela:1988cp}, 
allowing us to evaluate nonperturbative hadronic matrix elements of 
local operators that enter the expansion in (\ref{eq_daexpansion1}) 
in a fully controllable fashion, at least in principle.
 
In this work we report on the calculation of the first few moments 
of the leading-twist nucleon DA and also the normalization of the 
next-to-leading (twist-four) DAs~\cite{Braun:2000kw} using two 
dynamical flavors of clover fermions. The reason why we also consider 
higher-twist DAs is that they enter the calculation of the 
helicity-violating Pauli form factor of the nucleon in 
perturbative QCD~\cite{Belitsky:2002kj} and also the calculation of the
soft (end-point) corrections to the form factors in the framework of 
the light-cone sum rule approach~\cite{Braun:2001tj,Lenz:2003tq}. 
Their knowledge is imperative for a QCD description of exclusive 
reactions in the JLAB energy range. 
It turns out that the same matrix elements are responsible for 
proton decay in Grand Unified Theories (GUTs), so they are 
also interesting in a broader physics context.     
A short presentation of our main results has already been given in
Ref.~\cite{Gockeler:2008xv,Warkentin:2008iu}.

The paper is organized as follows. Section~\ref{sec_basics} contains 
a brief review of the general framework and definitions of the 
specific quantities that 
will be calculated. We focus on the relations to local
matrix elements including those that are relevant for proton decay.

In Section~\ref{sec_calc} we explain the lattice approach to the 
calculation of the matrix elements. The advantages of this method 
come at the cost of reduced symmetry due to the discretization of space-time. 
This leads to additional (unwanted) operator mixing as compared to the
continuum, which has to be reduced as much as possible by a suitable choice 
of the operator basis. 
In particular, mixing with lower-dimensional operators is dangerous.
The theoretical basis for the corresponding analysis is the 
classification of operators according to irreducible representations 
of the relevant lattice symmetry group.  
For quark-antiquark operators such a classification has been worked out in
Ref.~\cite{Gockeler:1996mu}, while the analogous classification for the
three-quark operators needed here is treated in
Refs.~\cite{Gockeler:2007qs,Kaltenbrunner:2008pb}.

Section~\ref{sec_results} is devoted to the presentation of  
the numerical results for the matrix elements.
We apply  two different methods  to analyze the data. The first one,
which we refer to as unconstrained,   
is used to determine the normalization constants and to check the
consistency of our results for higher moments. In the second method we use
the momentum conservation as an additional constraint. 
This allows us to improve the accuracy of our results for the higher moments.

In Section~\ref{sec:modelda} we construct a model for the leading-twist DA, 
presenting our results in form of the canonical expansion 
Eq.~(\ref{eq_daexpansion1}),
and compare  it with other models in the literature. 
The final  Section~\ref{sec:summary} is reserved for a summary and conclusions.

Some further technical details are presented in the Appendices, 
in particular the relations between the local operators relevant 
for leading-twist DAs of spin-1/2 baryons and the irreducible 
three-quark operators. We also present here the bare lattice results.

\section{General Framework \label{sec_basics}}

\subsection{Leading twist}

The leading-twist proton DA can be 
defined~\cite{Henriques:1975uh,Chernyak:1983ej} 
from a matrix element of a gauge-invariant nonlocal
three-quark operator:
\begin{eqnarray}
\lefteqn{ \langle 0 | u_\alpha^{a'}(z_1)u_\beta^{b'}(z_2)d_\gamma^{c'}(z_3)
\, U_{a'a}(z_1,z_0)U_{b'b}(z_2,z_0)U_{b'b}(z_3,z_0)
\epsilon^{abc}|p \rangle = }
\nonumber \\
&& =\frac{f_N}{4}\left \{(\not\! pC)_{\alpha\beta}(\gamma_5 N)_\gamma V(z_i p)+
(\not\! p \gamma_5 C)_{\alpha\beta} N_\gamma A(z_i p)
+\left (i\sigma_{\mu\nu}p^\nu C\right)_{\alpha\beta} (\gamma_\mu\gamma_5 N)_\gamma T(z_ip)
\right \}\,+\ldots
\label{eq_nonlocal}
\end{eqnarray}
Here $\sigma_{\mu\nu}=\frac{i}{2}[\gamma_\mu,\gamma_\nu]$,
$C$ is the charge conjugation matrix, $| p \rangle $
is a proton state with momentum $p$,
and $N$ is the proton spinor; ellipses stand for the higher-twist
constributions.
All interquark separations are assumed to be light-like,
e.g., $u(z_1)$ denotes the $u$-quark field at the space point $z_1n$
with $n^2=0$, and $U(z_n,z_0)$ denotes the non-Abelian
phase factor (light-like Wilson line)
\begin{equation}
   U(z_n,z_0) \equiv {\rm P\ exp\,}
   \left[ig\int_0^1\!dt\, (z_n-z_0)\,n_\mu A^\mu
   (tz_n+(1-t)z_0)\right].
\end{equation}
Because of the light-cone kinematics, the matrix element does not
depend on $z_0$ and the phase factors can be eliminated by choosing
a suitable gauge. 

The  invariant functions $V$, $A$ and $T$ can be presented in the form
\begin{equation}
V(z_i p)\equiv
 \int\![dx]\, \exp \Big[-i\sum x_i z_i (p\cdot n)\Big] V(x_i),
\label{xp}
\end{equation} 
and similarly for $A$ and $T$, where the integration measure is defined as
\begin{equation}
 \int\![dx] \equiv \int_0^1\! dx_1\,dx_2\,dx_3\,\delta
     (1-x_1-x_2-x_3)\,.
\end{equation}
The variables $x_i$ have the meaning of
the longitudinal momentum fractions carried by the three quarks in
the proton, $0\le x_i\le 1$ and $\sum x_i=1$.

The identity of the two $u$-quarks in (\ref{eq_nonlocal}) implies  
the following symmetry properties~\cite{Chernyak:1983ej} 
\begin{equation}
V(x_1,x_2,x_3)=V(x_2,x_1,x_3), \ \ \ \ A(x_1,x_2,x_3)=-A(x_2,x_1,x_3), 
\  \ \ \ \ T(x_1,x_2,x_3)=T(x_2,x_1,x_3).
\label{12nuc}
\end{equation}
In addition, the requirement that the proton has isospin $1/2$
yields the relation 
\begin{equation}
 2T(x_1,x_2,x_3)=[V-A](x_1,x_3,x_2)-[V-A](x_2,x_3,x_1)
\label{eq_isospinrelation}
\end{equation}   
so that all three invariant functions can be expressed in terms of a 
single DA $\varphi$  defined as
\begin{eqnarray}
 \varphi(x_1,x_2,x_3) = V(x_1,x_2,x_3)-A(x_1,x_2,x_3)\,.
\label{eq_nucleonDA}
\end{eqnarray}
The normalization convention is such that 
\begin{equation}
      \int[dx]\, \varphi(x_1,x_2,x_3) = 1\,.
\end{equation}
The definition in (\ref{eq_nonlocal}) is equivalent to the following
form of the proton state~\cite{Chernyak:1983ej,Chernyak:1987nu}
\begin{equation}
| p,\uparrow \rangle = f_N \int \frac{[dx]\, \varphi(x_i)}
                                    {2\sqrt{24x_1x_2x_3}}
\left \{
|u^\uparrow(x_1)u^\downarrow(x_2) d^\uparrow(x_3)\rangle
-
|u^\uparrow(x_1)d^\downarrow(x_2) u^\uparrow(x_3)\rangle \right \},
\end{equation}
where the arrows indicate the helicities and
the standard relativistic normalization for the states and
Dirac spinors is implied.

Moments of DAs are defined as
   \begin{equation}
      V^{lmn}=\int_0^1 [\mathrm dx]\; x_1^l x_2^m x_3^n\; V(x_1, x_2, x_3)
   \end{equation} 
and similarly for the other functions. They can be related to matrix elements of the local operators
   \begin{align}
      \mathcal V_\tau^{\rho \bar l \bar m \bar n }(0)\equiv&
      \mathcal V_\tau^{\rho (\lambda_1\cdots\lambda_l) (\mu_1\cdots\mu_m) (\nu_1\cdots\nu_n) }(0)
       \nonumber\\
              = & \epsilon^{a b c} \;
                                       \left[\imi^l D^{\lambda_1}\dots D^{\lambda_l}u(0)\right]^a_\alpha 
                                          \;     (C\gamma^\rho)_{\alpha\beta}       \;
                                       \left[\imi^m D^{\mu_1}\dots D^{\mu_m} u(0) \right]^b_\beta \;
                                       \left[\imi^n D^{\nu_1}\dots D^{\nu_n} (\gamma_5 d(0) ) \right]^c_\tau,
      \label{eq:vop}\\
      \mathcal A_\tau^{\rho \bar l \bar m \bar n}(0)\equiv&
      \mathcal A_\tau^{\rho (\lambda_1\cdots\lambda_l) (\mu_1\cdots\mu_m) (\nu_1\cdots\nu_n) }(0)
       \nonumber\\
              =&\epsilon^{a b c} \;
                                       \left[(\imi^l D^{\lambda_1}\dots D^{\lambda_l} u(0)\right]_\alpha^a
                                          \;     (C\gamma^\rho\gamma_5)_{\alpha\beta}      \;
                                       \left[\imi^m D^{\mu_1}\dots D^{\mu_m} u(0)\right]_\beta^b \;
                                       \left[\imi^n D^{\nu_1}\dots D^{\nu_n}  d(0)\right]^c_\tau ,
      \label{eq:aop}\\
      \mathcal T_\tau^{\rho \bar l \bar m \bar n}(0)\equiv&
       \mathcal T_\tau^{\rho (\lambda_1\cdots\lambda_l) (\mu_1\cdots\mu_m) (\nu_1\cdots\nu_n) }(0)
       \nonumber\\
              =&\epsilon^{a b c} \;
                                       [\imi^l D^{\lambda_1}\dots D^{\lambda_l} u(0)]_\alpha^a
                                          \;     \left(C(-i\sigma^{\xi\rho})\right)_{\alpha\beta}        \;
                                       [\imi^m D^{\mu_1}\dots D^{\mu_m} u(0)]_\beta^b    \;
                                       [\imi^n D^{\nu_1}\dots D^{\nu_n} (\gamma_\xi\gamma_5 d(0)) ]^c_\tau
      \label{eq:top}
  \end{align}
by
   \begin{align}
      \label{eq_opmatrix1}
      P_{LTW}\; \langle 0\vert \mathcal V_\tau^{\rho \bar l \bar m \bar n }(0) \vert p\rangle &=
         - f_N V^{l m n} p^\rho p^{\bar l} p^{\bar m} p^{\bar n}  N_\tau (p),\\
      P_{LTW}\; \langle 0\vert \mathcal A_\tau^{\rho \bar l \bar m \bar n }(0) \vert p\rangle &=
         - f_N A^{l m n} p^\rho p^{\bar l} p^{\bar m} p^{\bar n} N_\tau (p),\\
      P_{LTW}\; \langle 0\vert \mathcal T_\tau^{\rho \bar l \bar m \bar n }(0) \vert p\rangle &=
         2 f_N T^{l m n} p^\rho p^{\bar l} p^{\bar m} p^{\bar n} N_\tau (p).
      \label{eq_opmatrix3}
   \end{align}
In the following we refer to these local operators as DA operators 
in order to distinguish them from three-quark operators with a general
spinor index structure.
The multi-index $ \bar l \bar m \bar n $ with 
$\bar l\equiv\lambda_1\dots\lambda_l$ (and similarly 
for $\bar m$ and $\bar n$) denotes the Lorentz structure given by 
the covariant derivatives $D_{\mu} =\partial_\mu -i \mathrm g A_\mu$ 
on the right-hand side of
Eqs.~\eqref{eq:vop}-\eqref{eq:top}. The indices $l,m,n$ (without bars) are the
total number of derivatives acting on the first, second and third quark,
respectively. A certain moment, e.g.,  $V^{lmn}$, is related to several
operators $V_\tau^{\rho \bar l \bar m \bar n }$ which differ only by their
Lorentz indices.  Therefore the moments $V^{lmn},\,A^{lmn},\,T^{lmn}$ on the
right-hand side of Eqs.~(\ref{eq_opmatrix1})-(\ref{eq_opmatrix3}) can be
calculated from different operators with same number of derivatives acting on
the quark fields. The index $\rho$ corresponds to the uncontracted 
Lorentz index of the gamma matrices in the operators. The leading-twist 
projection, $P_{LTW}$, can be achieved, e.g., by symmetrization in 
Lorentz indices and subtraction of traces. Our approach for handling 
the reduced symmetry of the discretized space-time properly is described in
Section~\ref{sec_calc}.

The symmetry relations (\ref{12nuc}) are translated into
similar relations for the moments:
   \begin{equation}
      V^{lmn}=V^{mln},\quad A^{lmn}=-A^{mln},\quad T^{lmn}=T^{mln}.
   \end{equation}
For further use we define the combination
   \begin{equation}
   \phi^{lmn}=\frac{1}{3}(V^{lmn}-A^{lmn}+2T^{lnm})\,.
   \end{equation} 
Taking into account the isospin relation (\ref{eq_isospinrelation}), 
the moments of  $V,\,A,\,T$ can be restored from the moments of $\phi$ by
   \begin{align}
   T^{lmn}=&\frac{1}{2}(\phi^{lnm}+\phi^{mnl}),\\
   V^{l m n}=&\frac{1}{2} \left(2 \phi ^{l m n}+2 \phi ^{m l n}-\phi ^{n l m}-\phi^{n m l}\right),\\ 
   A^{l m n}=&\frac{1}{2} \left(-2 \phi ^{l m n}+2 \phi ^{m l n}-\phi^{n l m}+\phi^{n m l}\right).
   \end{align}
The conventional proton DA $\varphi(x_i)$ (\ref{eq_nucleonDA}) is given in terms of $\phi(x_i)$ as  
   \begin{equation}
   \varphi(x_1,x_2,x_3) =  2 \phi(x_1,x_2,x_3) -   \phi(x_3,x_2,x_1)\,, \qquad 
   \varphi^{lmn}=2\phi^{lmn}-\phi^{nml}.
   \end{equation} 
Due to momentum conservation ($x_1+x_2+x_3=1$) there are additional relations 
between lower and higher moments:
\begin{equation}
   \phi^{lmn}=\phi^{(l+1)mn}+\phi^{l(m+1)n}+\phi^{lm(n+1)}. \label{eq:sumrule}
\end{equation}
In particular this implies
\begin{equation}
   1=\phi^{000}=\phi^{100}+\phi^{010}+\phi^{001}=\phi^{200}+\phi^{020}+\phi^{002}+2(\phi^{011}+\phi^{101}+\phi^{110})=\dots
     \label{eq:onesum}
\end{equation}

\subsection{Next-to-leading twist operators and proton decay}

In general, there exist three independent next-to-leading (twist-four) 
three-quark DAs, cf.~Ref.~\cite{Braun:2000kw}. In this work we only 
consider their normalization, which is related to
the contributions of local operators without derivatives. Thus the 
problem is simplified considerably since the general Lorentz 
decomposition of the relevant matrix
element involves only four structures:
   \begin{equation}
      \begin{split}
      4\langle 0 \vert \epsilon^{abc} 
               u^a_\alpha(0) u^b_\beta(0) d^c_\gamma(0) 
       \vert p \rangle=
       &V_1^0 (\slashed{p} C)_{\alpha\beta} (\gamma_5 N)_\gamma+V_3^0 m_N (\gamma_\mu C)_{\alpha\beta} (\gamma^\mu\gamma_5 N)_\gamma\\
       &+T_1^0 (p^\nu i \sigma_{\mu\nu} C)_{\alpha\beta} (\gamma^\mu\gamma_5 N)_\gamma+ T_3^0 m_N (\sigma_{\mu\nu} C)_{\alpha\beta} (\sigma^{\mu\nu}\gamma_5 N)_\gamma\; ,
       \end{split}
   \end{equation} 
where $m_N$ is the nucleon mass and we have used the same notation as 
in~\cite{Braun:2000kw}. The leading-twist-three constants $V_1^0$ 
and $T_1^0$ correspond
to $f_N V^{000}$ and $f_N T^{000}$ in our notation, Eqs.~\eqref{eq_opmatrix1}
and \eqref{eq_opmatrix3}, and are equal. The two
additional constants, $V^0_3$ and $T^0_3$, correspond to 
subleading twist-four contributions. 
The combinations $\lambda_1=V_1^0-4V_3^0$ and $\lambda_2=6(V_1^0-4T_3^0)$ are
often arising in QCD sum rule calculations. They describe the nucleon 
coupling to the two independent local operators 
\begin{align}
      \mathcal L_\tau (0)&=
      \epsilon^{a b c}  \left[ {u^a}^T(0)  C\gamma^\rho u^b(0) \right ]\times (\gamma_5 \gamma_\rho d^c(0))_\tau,
      \label{eq:lambda1op}   \\
      \mathcal M_\tau (0)&=
      \epsilon^{a b c}  \left[ {u^a}^T(0)  C\sigma^{\mu\nu} u^b(0) \right ]\times (\gamma_5 \sigma_{\mu\nu} d^c(0))_\tau\,,
      \label{eq:lambda2op}
   \end{align}
which have been introduced in~\cite{Ioffe:1981kw,Chung:1981cc}, respectively.
Their matrix elements are given by
   \begin{align}
      \langle 0 \vert  \mathcal L_\tau (0) \vert p\rangle &=\lambda_1 m_N N_\tau, \label{eq_lambda1matrix}
\\
      \langle 0 \vert  \mathcal M_\tau (0) \vert p\rangle &=\lambda_2 m_N N_\tau.\label{eq_lambda2matrix}
   \end{align}
Separating the components of different helicity, one can write
\begin{align}
 \mathcal L_\tau=& 4\left(\gamma_R \, \mathcal U^{L}-\gamma_L\, \mathcal
U^{R}\right)_\tau\,,  \label{eq:Ldeco} \\
 \mathcal M_\tau=& 8\left(\gamma_R \, \mathcal U^{R}-\gamma_L\, \mathcal
U^{L}\right)_\tau\,,  \label{eq:Mdeco} 
\end{align}
where $\gamma_L=(1-\gamma_5)/2$ , $\gamma_R=(1+\gamma_5)/2$ are the left- and
right-handed projectors and
\begin{equation}
 \mathcal U^{L/R}_\tau= \epsilon^{abc} u_\tau^a \left[{(\gamma_{L/R} u^b)}^T C
\gamma_{L/R} d^c\right]  \,.
 \label{eq:uop}
\end{equation}
The Fierz identity implies
\begin{align}
 \epsilon^{abc}\left[{u^a}^T(0)C\gamma^\mu u^b(0)\right] \left(\gamma_5
\gamma_\mu d^c(0)\right)_\tau
     &=
 2 \epsilon^{abc}\left(
     -\left[{u^a}^T(0) C\gamma_5 d^b(0)\right] u^c(0)_\tau + \left[{u^a}^T(0) C
d^b(0)\right] \left( \gamma_5 u^c(0)\right)_\tau
 \right),\\
\epsilon^{abc}\left[{u^a}^T(0)C\sigma^{\mu\nu} u^b(0)\right] \left(\gamma_5
\sigma_{\mu\nu} d^c(0)\right)_\tau
     &=
 4 \epsilon^{abc}\left(
     \left[{u^a}^T(0) C\gamma_5 d^b(0)\right] u^c(0)_\tau + \left[{u^a}^T(0) C
d^b(0)\right] \left( \gamma_5 u^c(0)\right)_\tau
 \right).
\end{align}
Thus we get
\begin{equation}
 m_N(2\lambda_1+\lambda_2)N(p)=8 \langle 0\vert \epsilon^{abc} \left({u^a}^T C
d^b\right) \gamma_5 u^c \vert p\rangle \,,
 \label{eq:lambdasum}
\end{equation}
where, as it can be shown,  the matrix element on the right-hand side vanishes
in the nonrelativistic limit.

The operators (\ref{eq:lambda1op}) and (\ref{eq:lambda2op}) appear
also in the low-energy effective action of generic GUT models, and 
their matrix elements 
$\langle\pi |{\mathcal L}|p\rangle$ and
$\langle\pi |{\mathcal M}|p\rangle$ give rise to proton decay. 
These matrix elements, in turn, can be related to the constants defined in 
(\ref{eq_lambda1matrix}), (\ref{eq_lambda2matrix}), 
using soft pion theorems or, what is the same, leading order in 
chiral perturbation theory~\cite{Tomozawa:1980rc, Wise:1980ch, Claudson:1981gh,
Berezinsky:1981qb, Brodsky:1983st}.

\begin{figure}[ht]
   \includegraphics[clip,width=0.7\textwidth]{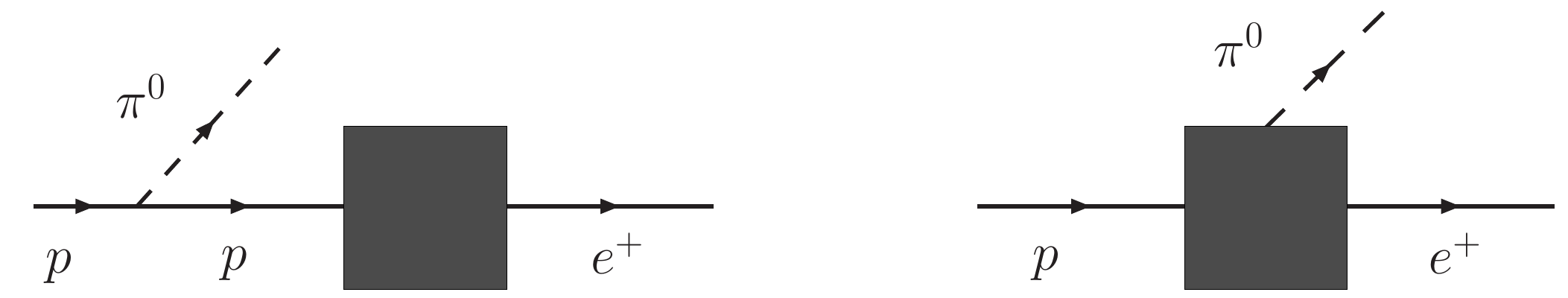}
   \caption{Diagrams contributing to the nucleon decay amplitude $p\rightarrow\pi^0+e^+$.\label{fig_pdecay}}
\end{figure}

To this end one introduces two low-energy constants $\alpha$ and $\beta$  which extend the usual
three-flavor baryon chiral Lagrangian. They are defined by
   \begin{align}
	\langle 0 \vert  (\gamma_L \mathcal U^R)_\tau (0) \vert p\rangle &= - \alpha (\gamma_L N)_\tau,
	& 
	\langle 0 \vert  (\gamma_R \mathcal U^L)_\tau (0) \vert p\rangle &= \alpha (\gamma_R N)_\tau, 
	\\
      	\langle 0 \vert  (\gamma_L \mathcal U^L)_\tau (0) \vert p\rangle &= - \beta  (\gamma_L N)_\tau,
	&
	\langle 0 \vert  (\gamma_R \mathcal U^R)_\tau (0) \vert p\rangle &= \beta  (\gamma_R N)_\tau.
   \end{align}
Because of \eqref{eq:Ldeco}, \eqref{eq:Mdeco} one obtains 
$\alpha = m_N \lambda_1/4$ and $\beta = m_N \lambda_2/8$. 
The knowledge of these two constants allows one to estimate 
nucleon-to-pion decay matrix elements. Using the notation of 
Ref.~\cite{Aoki:2006ib} the relevant factors in the decay amplitude for
the proton to $\pi^0$ decay (cf. Fig.~\ref{fig_pdecay}) have the form
   \begin{align}
      W_0^{R L}(p\rightarrow \pi^0)=& \frac{\alpha}{\sqrt{2}f} (1+g_A),\\
      W_0^{L L}(p\rightarrow \pi^0)=& \frac{\beta}{\sqrt{2}f}  (1+g_A),
   \end{align}
where $f$ is the tree level pion decay constant normalized such that the 
experimental value is $f_\pi\simeq 131\;\textrm{MeV}$ and $g_A$ is the 
axial charge.

\section{Details of the lattice calculation \label{sec_calc}}

In this section we discuss the techniques used and the details of the lattice
calculation. From now on  we work in Euclidean space. In order to define the
Euclidean counterparts of the operators Eqs.~\eqref{eq:vop}-\eqref{eq:top}, 
\eqref{eq:lambda1op}-\eqref{eq:lambda2op} and \eqref{eq:uop} we interpret the
Dirac matrices and coordinates as being Euclidean. For our Euclidean Dirac
matrices see Appendix~\ref{app:weyl}. The expressions on the right-hand-side of
Eqs.~\eqref{eq:vop}-\eqref{eq:top}, \eqref{eq:lambda1op}-\eqref{eq:lambda2op}
are then modified accordingly. In the first part of this Section we summarize
the general features of our approach. The following parts contain the
description of the calculation of matrix elements relevant for leading and
next-to-leading twist DAs.

\subsection{General features}
To be as flexible as possible in our calculation we have adopted 
a two-stage approach in the evaluation of the correlators. In the first 
step we have calculated correlators of the form
   \begin{equation}
      C^{\bar l \bar m \bar n}_{\alpha\beta\gamma \tau}=\langle \epsilon^{abc}
		[D_{\lambda_1}\dots D_{\lambda_l} u(x)]^a_\alpha 
		[D_{\mu_1}\dots D_{\mu_m} u(x)]^b_\beta 
		[D_{\nu_1}\dots D_{\nu_n} d(x)]^c_\gamma  \bar{\mathcal N}(y)_\tau \rangle,
	\label{eq_gen3qop}
   \end{equation}
with $l+m+n\leq 2$.  As interpolating operator for the proton we have used
\begin{equation}
 \mathcal N_\tau=\epsilon^{abc} \left[{u^a}^T C \gamma_5 d^b\right]  u^c_\tau\,.
\end{equation} 
Due to the presence of two $u$-quarks in the three-quark operator, $C^{\bar
m \bar l \bar n}_{\alpha\beta\gamma \tau}$ can be reconstructed from $C^{\bar l
\bar m \bar n}_{\alpha\beta\gamma \tau}$ by an appropriate interchange of Dirac
indices.

In the second step the general three-quark operator from 
Eq.~\eqref{eq_gen3qop} 
was used to calculate the matrix elements for the different quantities we
discussed before. The general form of the correlation functions we compute at
this stage reads after projection onto momentum $\vec p$:
\begin{equation}
\langle \mathcal O_\tau(t,\vec p) \bar {\mathcal N}_{\tau^\prime}(0,\vec p)\rangle =
 \frac{\sqrt{Z_N(\vec p)}}{2 E(\vec p)}   
\sum_s
	\langle 0 \vert \mathcal O_\tau(0)  \vert p, s \rangle 
	\bar N_{\tau^\prime}( p,s) \exp \left( -E(\vec p) t\right).
	\label{eq:gencorr}
\end{equation}
Here contributions of excited states have been neglected and the dependence of 
the nucleon states and spinors on the spin vector $s$ has been made explicit. 
For the energy $E(\vec p)$ we use the continuum expression 
$E(\vec p)=\sqrt{m_N^2+\vec p{\,}^2}$. We have checked that this dispersion
relation is fulfilled well within errors (see, e.g.,
Fig.~\ref{fig:effmass}), so we had to fit only the mass in the
exponential. The correlator in Eq.~\eqref{eq:gencorr} can be directly
constructed from the general correlation function \eqref{eq_gen3qop}. 
The matrix element on the right-hand side is the quantity we want 
to determine. Thus we
have also to calculate the normalization constant $Z_N(\vec p)$, which can be
extracted from the usual two-point nucleon correlator
   \begin{equation}
      C_N(\vec p)\equiv\left(\gamma_+ \right)_{\tau^\prime\tau} \langle \mathcal N_\tau(t,\vec p) \bar{\mathcal N}_{\tau^\prime}(0, \vec p)\rangle = 
       Z_N(\vec p ) \frac{m_N+E(\vec p)}{E(\vec p)} \exp{(-E(\vec p)t)}
	\label{eq_nncorr}
   \end{equation} 
with the positive parity projection $\gamma_+=(1+\gamma_4)/2$.
In the evaluation of the correlator in Eq.~(\ref{eq_gen3qop}) the overlap of
the nucleon interpolator with the nucleon state is improved by Jacobi smearing
at the source while the sink is not smeared since we want to evaluate local
matrix elements. Thus the nucleon correlator in Eq.~(\ref{eq_nncorr}) 
cannot be extracted from the general three-quark nucleon correlator 
\eqref{eq_gen3qop} but must be computed separately with Jacobi 
smeared sink and source.

The normalization constant $Z_N(\vec p)$ could be removed by 
considering the ratio
   \begin{equation}
      \frac{\left((\gamma_+)_{\tau^\prime\tau}\langle \mathcal O_\tau(t) \bar{\mathcal N}_{\tau^\prime}(0)\rangle\right)^2}
	   {(\gamma_+)_{\tau^\prime\tau}\langle \mathcal N_\tau(t) \bar{\mathcal N}_{\tau^\prime}(0)\rangle}.
   \end{equation} 
However, as we will see later, the location of the effective mass plateaus is
different for the two correlators, presumably due to the different smearings on
the sink, spoiling this simple approach. Thus instead of calculating the ratio
we perform a correlated fit to the two correlators in the range of the
corresponding effective mass plateaus.

Up to now we did not take into account that our calculations are performed on a
space-time lattice. This leads to reduced symmetry compared to the continuum.
Due to this symmetry reduction we expect additional operator mixings which are
not present in the continuum. In particular, we can have mixing with
lower-dimensional operators. Thus a systematic analysis and careful choice of
the operators used is mandatory. In \cite{Kaltenbrunner:2008pb} a complete
classification with respect to the spinorial extension of the hypercubic group
$H(4)$ for all three-quark operators without derivatives is presented. For
operators with one and two derivatives the classification is worked out for the
leading-twist case. These results enable us to derive operators 
with ``good'' mixing properties, good in the sense that they do not mix with
lower-dimensional operators. They belong to definite irreducible
representations of the spinorial extension of $H(4)$ and are most easily
constructed in the Weyl representation of the Dirac matrices. Therefore we 
also work in this representation.

\begin{table}[t]
\renewcommand{\arraystretch}{1.5}
   \begin{tabular}{|c||c|c|c|}
   \hline
                              &  $d=9/2$  ($0$ derivatives)  &   $d=11/2$ ($1$ derivative)  &   $d=13/2$  ($2$ derivatives)\\ \hline \hline
      $\tau^{\underline{4}}_1$&  
                                 $\mathcal B_{1,i}^{(0)}$, $\mathcal B_{2,i}^{(0)}$, $\mathcal B_{3,i}^{(0)}$, $\mathcal B_{4,i}^{(0)}$, $\mathcal B_{5,i}^{(0)}$ &  &  $\mathcal B_{1,i}^{(2)}$, $\mathcal B_{2,i}^{(2)}$,  $\mathcal B_{3,i}^{(2)}$ \\ \hline
      $\tau^{\underline{4}}_2$&     &     & $\mathcal B_{4,i}^{(2)}$, $\mathcal B_{5,i}^{(2)}$,  $\mathcal B_{6,i}^{(2)}$ \\ \hline
      $\tau^{\underline{8}}$&   $\mathcal B_{6,i}^{(0)}$  &     
                                $\mathcal B_{1,i}^{(1)}$  & 
                                $\mathcal B_{7,i}^{(2)}$, $\mathcal B_{8,i}^{(2)}$,  $\mathcal B_{9,i}^{(2)}$ \\ \hline
      $\tau^{\underline{12}}_1$& $\mathcal B_{7,i}^{(0)}$,  $\mathcal B_{8,i}^{(0)}$,  $\mathcal B_{9,i}^{(0)}$ &     
                                $\mathcal B_{2,i}^{(1)}$,   $\mathcal B_{3,i}^{(1)}$, $\mathcal B_{4,i}^{(1)}$  &
                                $\mathcal B_{10,i}^{(2)}$, $\mathcal B_{11,i}^{(2)}$, $\mathcal B_{12,i}^{(2)}$, $\mathcal B_{13,i}^{(2)}$\\ \hline
      $\tau^{\underline{12}}_2$&     &     
                                $\mathcal B_{5,i}^{(1)}$,   $\mathcal B_{6,i}^{(1)}$, $\mathcal B_{7,i}^{(1)}$ , $\mathcal B_{8,i}^{(1)}$ &
                                $\mathcal B_{14,i}^{(2)}$, $\mathcal B_{15,i}^{(2)}$, $\mathcal B_{16,i}^{(2)}$, $\mathcal B_{17,i}^{(2)}$,  $\mathcal B_{18,i}^{(2)}$\\ \hline
   \end{tabular}
   \caption{\label{tab_irredrepr} Overview of irreducibly transforming
multiplets of three-quark operators sorted by their mass dimension (number of
derivatives) taken from \cite{Kaltenbrunner:2008pb} with a notation adapted to
our needs.  Since for the classification it is not important on which quarks the
derivatives act, only the sum $l+m+n$ is given as a superscript. The subscript
gives the numbering of the operators according to the numbering convention in
\cite{Kaltenbrunner:2008pb}. The first number corresponds to the lower index of
\cite{Kaltenbrunner:2008pb} while the second number corresponds to the upper
index in \cite{Kaltenbrunner:2008pb} labelling different operators within one
multiplet (cf. Table~4.1 in \cite{Kaltenbrunner:2008pb}). In the first column we
give also the representations in the notation of  \cite{Kaltenbrunner:2008pb}
where the superscript denotes the dimension.
}
\end{table}

In Table~\ref{tab_irredrepr} we give an overview of the irreducible multiplets
of operators taken from Table~4.1 in \cite{Kaltenbrunner:2008pb}, with a
modified notation adapted to our needs, e.g., operator $B_{1,i}^{(2)}$
corresponds to $\mathcal O_{DD1}^{(i)}$ in \cite{Kaltenbrunner:2008pb} and
similarly for the others. The next-to-leading twist
operators (\ref{eq:lambda1op}) and (\ref{eq:lambda2op})
lie completely within the $\tau^{\underline{4}}_1$ representation with mass
dimension $9/2$. The operators relevant for the leading-twist DAs
belong to other multiplets.  As operators without derivatives in the 
$\tau^{\underline{8}}$ representation do not have an overlap with the nucleon,
the relevant operators with ``good'' mixing properties lie in 
$\tau^{\underline{12}}_1$, $\tau^{\underline{12}}_2$ and
$\tau^{\underline{4}}_2$ for zero, one and two derivatives, respectively. 
Rewriting these irreducible operators in terms of the DA operators defined in 
\eqref{eq:vop}-\eqref{eq:top} allows us to choose those that are 
suited for lattice calculations. The ensuing relations for 
leading-twist spin-1/2 baryon DAs are summarized in Appendix~\ref{app:mom}. 
In the following we give some details for these operators.

Initially, the irreducible operators in \cite{Kaltenbrunner:2008pb} 
have a general flavor content. Considering the case of two derivatives 
as an example we have operators of the type
\begin{equation}
 \Gamma_{\mu\nu}^{\alpha\beta\gamma}D_\mu D_\nu \epsilon^{abc} f^a_{\alpha} g^b_{\beta} h^c_{\gamma}\, ,
\end{equation}
where $\Gamma_{\mu\nu}^{\alpha\beta\gamma}$ is a tensor projecting the 
operator to a certain irreducible representation. As it is not important 
for the construction of irreducibly transforming operators on which 
of the quarks the derivatives act, the different possibilities fall 
into the same irreducible representation. The proton operators are 
then recovered by the identification
\begin{equation}
f\rightarrow u,\qquad g\rightarrow u, \qquad h\rightarrow d,
\end{equation}
and subsequent projection onto isospin $1/2$,
which is done by combining properly different multiplets.
This procedure differs somewhat from the approach adopted in 
Ref. \cite{Kaltenbrunner:2008pb}, but it leads to equivalent results.

The operators used in our calculation have to be renormalized.
In \cite{Gockeler:2008we,PHDThomas:2008} the required renormalization matrices 
were calculated nonperturbatively on the lattice imposing an 
RI$^\prime$-MOM-like renormalization condition. Using  continuum 
perturbation theory and 
the renormalization group the results were converted to the
\msb\ scheme at a scale of $4\,\mathrm{GeV}^2$.  Note that in 
this procedure the mixing with ``total derivatives'' is automatically taken 
into account. 
The scale at which our renormalization condition is imposed is taken
to be $20\,\mathrm{GeV}^2$, and the systematic uncertainty is estimated
by varying this scale between $10\,\mathrm{GeV}^2$ and $40\,\mathrm{GeV}^2$.

\subsection{Moments of the leading-twist DA}

\subsubsection*{0th moment}

Using the representation $\tau^{\underline{12}}_1$ and the relations to 
the DA operators given in Appendix~\ref{app_mom0} we construct 
three quadruplets of operators with isospin $1/2$ 
from the twelve irreducible three-quark operators,
which can be used to calculate $f_N$:
   \begin{align}
      \mathcal O_{A,0}^{000}=&\frac{4}{3}\left(
      \begin{array}{rcl}
      -\mathcal{B}^{000}_{8,6}&+&\mathcal{B}^{000}_{9,6}\\
      \mathcal{B}^{000}_{8,1}&-&\mathcal{B}^{000}_{9,1}\\
      -\mathcal{B}^{000}_{8,12}&+&\mathcal{B}^{000}_{9,12}\\
      \mathcal{B}^{000}_{8,7}&-&\mathcal{B}^{000}_{9,7}\\
      \end{array}
      \right),&
      \mathcal O_{B,0}^{000}=&\frac{4}{3}\left(
      \begin{array}{rcl}
      -\mathcal{B}^{000}_{8,4}&+&\mathcal{B}^{000}_{9,4}\\
      \mathcal{B}^{000}_{8,3}&-&\mathcal{B}^{000}_{9,3}\\
      -\mathcal{B}^{000}_{8,10}&+&\mathcal{B}^{000}_{9,10}\\
      \mathcal{B}^{000}_{8,9}&-&\mathcal{B}^{000}_{9,9}\\
      \end{array}
      \right),&
      \mathcal O_{C,0}^{000}=&
		\frac{4 \sqrt{2}}{3}
	\left(
      \begin{array}{rcl}
      \mathcal{B}^{000}_{8,2}&-&\mathcal{B}^{000}_{9,2}\\
      -\mathcal{B}^{000}_{8,5}&+&\mathcal{B}^{000}_{9,5}\\
      \mathcal{B}^{000}_{8,8}&-&\mathcal{B}^{000}_{9,8}\\
      -\mathcal{B}^{000}_{8,11}&+&\mathcal{B}^{000}_{9,11}\\
      \end{array}
      \right).&
   \end{align}
The three-quark operators $\mathcal O$ on the left-hand side have 
also a Dirac index which we do not give explicitly here. 
The relations to the DA operators given in Appendix~\ref{app_mom0} yield then
   \begin{align}
      \langle 0|\mathcal O_{A,0}^{000} \vert  p\rangle&=f_N
      (i p_1\gamma_1-i p_2\gamma_2)N(p),\\
      \langle 0|\mathcal O_{B,0}^{000} \vert p\rangle&=f_N
      (i p_3\gamma_3+E(\vec p)\gamma_4)N(p),\\
      \langle 0|\mathcal O_{C,0}^{000} \vert p \rangle&=f_N
      (i p_1\gamma_1+i p_2\gamma_2- i p_3\gamma_3+E(\vec p)\gamma_4)N(p).
   \end{align} 
The operators $\mathcal O_{B,0}^{000}$ and $\mathcal O_{C,0}^{000}$ are most
suitable for our calculation since $\mathcal O_{A,0}^{000}$ would
require nonzero spatial momenta in the 1 or 2 direction, which would increase
the statistical noise.  Thus, in order to determine $f_N$, we evaluate
finally only the following two correlators at $\vec{p} =\vec 0$:
   \begin{align}
      C_{B,0}^{000}\equiv\langle \left(\gamma_4 \mathcal O_{B,0}^{000}(t,\vec
p)\right)_\tau 
		\left(\bar{\mathcal N}(0,\vec p)\right)_{\tau^\prime} \left(\gamma_+\right)_{\tau^\prime\tau}\rangle &=
               f_N \sqrt{Z_N(\vec p)}  
	\frac{E(\vec p)\left( m_N +E(\vec p)\right)+p_3^2}{E(\vec p)} \exp \left (-E(\vec p)t \right ),\\
     C_{C,0}^{000}\equiv \langle \left(\gamma_4 \mathcal O_{C,0}^{000}(t,\vec
p)\right)_\tau
	\left(\bar{\mathcal N}(0,\vec p)\right)_{\tau^\prime}\left(\gamma_+\right)_{\tau^\prime\tau}\rangle &=
               f_N \sqrt{Z_N(\vec p)}  \frac{E(\vec p)(m_N+E(\vec p))+p_1^2+p_2^2-p_3^2}{E(\vec p)} \exp \left (-E(\vec p)t \right ).
   \end{align}

\subsubsection*{1st moments}
We use the irreducible operators with one derivative from  
Appendix~\ref{app_mom1} to construct operators for the calculation 
of the first moments of the proton DA,
   \begin{align}
      \mathcal O_{A,1}^{lmn}=&\frac{4\sqrt{2}}{3}\left(
      \begin{array}{rcl}
      \mathcal{B}^{lmn}_{6,1} &-&\mathcal{B}^{lnm}_{7,1}\\
     -\mathcal{B}^{lmn}_{6,2} &+&\mathcal{B}^{lnm}_{7,2}\\
     -\mathcal{B}^{lmn}_{6,7} &+&\mathcal{B}^{lnm}_{7,7}\\
      \mathcal{B}^{lmn}_{6,8} &-&\mathcal{B}^{lnm}_{7,8}\\
      \end{array}
      \right),&
      \mathcal O_{B,1}^{lmn}=&\frac{4\sqrt{2}}{3}\left(
      \begin{array}{rcl}
      \mathcal{B}^{lmn}_{6,3} &-&\mathcal{B}^{lnm}_{7,3}\\
     -\mathcal{B}^{lmn}_{6,4} &+&\mathcal{B}^{lnm}_{7,4}\\
     -\mathcal{B}^{lmn}_{6,9} &+&\mathcal{B}^{lnm}_{7,9}\\
      \mathcal{B}^{lmn}_{6,10}&-&\mathcal{B}^{lnm}_{7,10}\\
      \end{array}
      \right),&
      \mathcal O_{C,1}^{lmn}=&\frac{4}{3}\left(
      \begin{array}{rcl}
      \mathcal{B}^{lmn}_{6,6}&-&\mathcal{B}^{lnm}_{7,6}\\
      \mathcal{B}^{lmn}_{6,5}&-&\mathcal{B}^{lnm}_{7,5}\\
     -\mathcal{B}^{lmn}_{6,12}&+&\mathcal{B}^{lnm}_{7,12}\\
     -\mathcal{B}^{lmn}_{6,11}&+&\mathcal{B}^{lnm}_{7,11}\\
      \end{array}
      \right), &
   \end{align}
where the the superscript $lmn$ with $l+m+n=1$ and nonnegative 
integers $l,m,n$ indicates on which fields the derivative acts. 
The matrix elements of these operators are then
   \begin{align}
      \langle 0|\mathcal O_{A,1}^{lmn} \vert p\rangle=& f_N\phi^{lmn}
      \left[
      (p_1 \gamma_1 -p_2 \gamma_2 )(i  p_3 \gamma_3 - E(\vec p) \gamma_4 ) - 2 i p_1 p_2\gamma_1\gamma_2
      \right] N(p), \\
      \langle 0|\mathcal O_{B,1}^{lmn} \vert p\rangle=& f_N\phi^{lmn}
      \left[
      (p_1 \gamma_1 + p_2 \gamma_2 )(i p_3 \gamma_3 +E(\vec p)\gamma_4)-2 p_3 E(\vec p)\gamma_3\gamma_4
      \right] N( p), \\
      \langle 0|\mathcal O_{C,1}^{lmn} \vert p\rangle=& f_N\phi^{lmn}
      (- p_1 \gamma_1 + p_2 \gamma_2 )(i p_3 \gamma_3 +E(\vec p)\gamma_4 )
       N( p),
   \end{align} 
where again a Dirac index is implied for the three-quark operators 
$\mathcal O$. Unlike the case of the $0$th moment all operators 
require at least one nonzero component of the spatial momentum. 
Hence using  all  operators available in this case we evaluate the correlators 
   \begin{align}
     C_{A,1}^{lmn}\equiv \langle \left( \gamma_4\gamma_1 \mathcal O_{A,1}^{lmn}(t,\vec p)\right)_\tau
	 \left( \bar{\mathcal N}(0,\vec p)\right)_{\tau^\prime}
	\left(\gamma_+\right)_{\tau^\prime\tau}
      \rangle=&
     -f_N \phi^{lmn} \sqrt{Z_N(\vec p)}\;  p_1 \frac{E(\vec p)(m_N +E(\vec p))+2p_2^2-p_3^2}{E(\vec p)}
	\exp \left (-E(\vec p)t \right ), \\      
      C_{B,1}^{lmn}\equiv \langle 
	\left(\gamma_4\gamma_1 \mathcal O_{B,1}^{lmn}(t,\vec p)\right)_\tau
	 \left(\bar{\mathcal N}(0,\vec p)\right)_{\tau^\prime}
	\left(\gamma_+\right)_{\tau^\prime\tau}
      \rangle=&
      \quad\: f_N \phi^{lmn} \sqrt{Z_N(\vec p)}\;  p_1 \frac{E(\vec p)(m_N +E(\vec p))+p_3^2}{E(\vec p)} \exp \left (-E(\vec p)t \right ), \\
     C_{C,1}^{lmn}\equiv \langle 
	\left(\gamma_4\gamma_1 \mathcal O_{C,1}^{lmn}(t,\vec p)\right)_\tau
	\left( \bar{\mathcal N}(0,\vec p)\right)_{\tau^\prime}
	\left(\gamma_+\right)_{\tau^\prime\tau}
       \rangle=&
      -f_N \phi^{lmn} \sqrt{Z_N(\vec p)}\;  p_1 \frac{E(\vec p)(m_N + E(\vec p))+p_3^2}{E(\vec p)} \exp \left (-E(\vec p)t \right )
   \end{align}
to determine the first moments $\phi^{100}$,  $\phi^{010}$ and  $\phi^{001}$.

\subsubsection*{2nd moments}
The calculation of the second moments requires the use of the 
four-dimensional irreducible representation 
$\tau^{\underline{4}}_2$ to avoid mixing with lower-dimensional operators. 
Unfortunately, this decreases also the number of possible operators. 
Using the irreducible three-quark operators with two derivatives and 
the relations to the DA operators from  Appendix~\ref{app_mom2} we construct
   \begin{equation}
      \mathcal O^{lmn}_2:=
      \frac{4}{3 \sqrt{3}}
      \left(
      \begin{array}{c}
      \mathcal{B}^{lnm}_{6,4}-\mathcal{B}^{lmn}_{5,4}\\
      \mathcal{B}^{lnm}_{6,3}-\mathcal{B}^{lmn}_{5,3}\\
      \mathcal{B}^{lnm}_{6,2}-\mathcal{B}^{lmn}_{5,2}\\
      \mathcal{B}^{lnm}_{6,1}-\mathcal{B}^{lmn}_{5,1}\\
      \end{array}
      \right)
   \end{equation}
where now $l+m+n=2$ with $l,m,n$ nonnegative integers. The corresponding 
matrix element is given by
   \begin{equation}
      \langle 0 \vert \mathcal O^{lmn}_2 \vert p\rangle=
          f_N  \phi^{lmn}
      \left[
      p_1 p_2\gamma_1\gamma_2\left(i p_3\gamma_3+E(\vec p)\gamma_4\right) +
      i p_3 E(\vec p) \gamma_3\gamma_4\left(i p_1\gamma_1 - i p_2\gamma_2\right)
      \right] N(p)
   \end{equation} 
and the second moments are determined from
   \begin{equation}
      C_2^{lmn}\equiv\langle 
	\left( \gamma_2 \gamma_3 \gamma_4 \mathcal O_2^{lmn}(t,\vec p) \right)_\tau
	\left( \bar{\mathcal N}(0,\vec p) \right)_{\tau^\prime}
	\left(\gamma_+\right)_{\tau^\prime\tau}
      \rangle=
      -f_N \phi^{lmn} \sqrt{Z_N(\vec p)}\; p_2 p_3 \frac{E(\vec p)(m_N+E(\vec p))+p_1^2}{E(\vec p)} \exp \left (-E(\vec p)t \right ).
   \end{equation}

\subsection{Next-to-leading twist DAs }
For the higher-twist DAs we consider only the operators 
without derivatives. If we write the operators in
Eqs.~\eqref{eq:lambda1op}-\eqref{eq:lambda2op} and \eqref{eq:uop} with
general flavor content,
\begin{align}
      \mathcal L_\tau (0)&=
      \epsilon^{a b c}  \left[ {f^a}^T(0)  C\gamma_\rho g^b(0) \right ]\times
(\gamma_5 \gamma_\rho h^c(0))_\tau\,,
      	\\
      \mathcal M_\tau (0)&=
      \epsilon^{a b c}  \left[ {f^a}^T(0)  C\sigma_{\mu\nu} g^b(0) \right
]\times (\gamma_5 \sigma_{\mu\nu} h^c(0))_\tau\,,
	\\
 \mathcal U^{L/R}_\tau(0) & = \epsilon^{abc} \left[{(\gamma_{L/R}
{g^b})^T(0)} C \gamma_{L/R} h^c(0)\right] \times  f_\tau^a(0) \,,
\end{align}
we can express them in terms of the irreducible three-quark operators 
as
   \begin{align}
      \mathcal L & = 
	\sqrt{8}
      \left(
         \begin{array}{c}
            \mathcal{B}^{lmn}_{3,1} + \mathcal{B}^{lmn}_{4,1}\\
            \mathcal{B}^{lmn}_{3,2} + \mathcal{B}^{lmn}_{4,2}\\
            \mathcal{B}^{lmn}_{3,3} + \mathcal{B}^{lmn}_{4,3}\\
            \mathcal{B}^{lmn}_{3,4} + \mathcal{B}^{lmn}_{4,4}
         \end{array}
      \right),&
            \mathcal M & = 
	\sqrt{96}
      \left(
         \begin{array}{c}
            \mathcal{B}^{lmn}_{2,1} \\
            \mathcal{B}^{lmn}_{2,2} \\
            \mathcal{B}^{lmn}_{2,3} \\
            \mathcal{B}^{lmn}_{2,4}
         \end{array}
      \right)
   \end{align}
and
   \begin{align}
     \gamma_R \mathcal  U^L - \gamma_L \mathcal U^R = &\sqrt{2}\,\left(
         \begin{array}{c}
            \mathcal{B}^{lmn}_{3,1} \\
            \mathcal{B}^{lmn}_{3,2} \\
            \mathcal{B}^{lmn}_{3,3} \\
            \mathcal{B}^{lmn}_{3,4}
         \end{array}
         \right),&
      \gamma_L \mathcal U^L - \gamma_R \mathcal U^R = &\sqrt{2/3}\, \left(
         \begin{array}{c}
             \mathcal{B}^{lmn}_{1,1} -\mathcal{B}^{lmn}_{2,1}\\
             \mathcal{B}^{lmn}_{1,2} -\mathcal{B}^{lmn}_{2,2}\\
             \mathcal{B}^{lmn}_{1,3} -\mathcal{B}^{lmn}_{2,3} \\
             \mathcal{B}^{lmn}_{1,4} -\mathcal{B}^{lmn}_{2,4}
         \end{array}
         \right)\,.
   \end{align}
After the identification $f\rightarrow u,\, g\rightarrow u, \, h\rightarrow d,$
we restore the proton operators in \eqref{eq:lambda1op}-\eqref{eq:lambda2op}
and \eqref{eq:uop}.

\section{Numerical results \label{sec_results}}

We have evaluated our correlators on the QCDSF/DIK configurations 
generated with two flavors of clover fermions at two different 
$\beta$ values summarized in Table~\ref{tab_latset}.
For $\beta=5.29$ we have used two different lattice sizes, 
$24^3\times 48$ and $16^3\times 32$, each at three different 
quark masses. For $\beta=5.40$ we have evaluated the correlators 
at five different quark masses on $24^3\times 48$ lattices. 
The lattice spacing has been set via the Sommer parameter 
$r_0=0.467\textrm{fm}$ \cite{Khan:2006de,Aubin:2004wf}. As far as 
possible we have also checked that the dependence of the final results on the 
fitting procedures discussed below is only very mild and the deviations are
consistent with the present statistical errors.

\subsection{General discussion}
\begin{table}[ht]
\begin{center}
     \begin{tabular}{|c|c|c|c|c|c|c|}                                                                          \hline
     $\beta$   & $\kappa$ & $m_\pi[\mathrm{GeV}]$  &        volume    & $a[\mathrm{fm}]$   & $L[\mathrm{fm}]$ \\ \hline \hline
     5.29     &  $0.1340$, $0.1350$,  $0.1359$& $1.411$, $1.029$, $0.587$ &  $16^3\times32$     & $0.08$  &  $1.28$\\ \hline 
     5.29     &   $0.1355$, $0.1359$, $0.1362$  & $0.800$, $0.587$, $0.383$&  $24^3\times48$ &  $0.08$ & $1.92$\\ \hline 
      
     5.40     &     $0.135$, $1356$, $0.1361$,  & $1.183$, $0.856$, $0.648$,& $24^3\times48$    &   $0.07$ &  $1.68$\\ 
          &  $0.13625$, $13640$  & $0.559$, $0.421$    &                 &            &\\\hline 
     \end{tabular}
\end{center}
\caption{The set of lattices used in our calculation. The scale was 
set via the Sommer parameter $r_0=0.467\,\mathrm{fm}$.
\label{tab_latset}}
\end{table}

As already anticipated we can reduce the noise by combining different momenta 
and/or different operators. However, calculating the general three-quark 
operator for many momenta turned out to be too expensive. Hence the general 
correlators (\ref{eq_gen3qop}), and therefore also the correlators for 
DA operators, were evaluated only for a minimal set of 
momenta. 

To extract the nucleon wave function normalization constant $f_N$ 
we have fitted the correlator
\begin{equation}
 C^{000}_0=\frac{1}{2}(C^{000}_{B,0}+C^{000}_{C,0})
\end{equation}
where we have averaged over the two possible correlators at $\vec p=\vec 0$. 
Similarly, for the first moments we have used
\begin{equation}
 C^{lmn}_1=\frac{1}{3}\left(C^{lmn}_{A,1}+C^{lmn}_{B,1}+C^{lmn}_{C,1}\right),
\end{equation}
with $l+m+n=1$ and $\vec p=(2\pi/L,0,0)$, where $L$ is the spatial extent of 
our lattice. For the second moment we have only one correlator, hence no
averaging is possible and we have evaluated it for $\vec p=(0,2\pi/L,2\pi/L)$.

To determine the normalization constant $Z_N(\vec p)$ we had also to evaluate 
the usual nucleon correlator. As the additional smearing on the sink introduces
additional noise, in particular for $\vec p\neq \vec 0$, we have improved the
signal by using different momenta in the nucleon correlator. For the 
$16^3\times 32$ lattices we have worked with
\begin{equation}
 C_N^1=\frac{1}{3}\left(C_N(2\pi/L,0,0)+C_N(0,2\pi/L,0)+C_N(0,0,2\pi/L)\right)\\
\end{equation}
and
\begin{equation}
C_N^2=\frac{1}{3}\left(C_N(0,2\pi/L,2\pi/L)+C_N(2\pi/L,0,2\pi/L)+C_N(2\pi/L,2\pi/L,0)\right), 
\end{equation}
while for the $24^3\times 48$ lattices we have used a larger number of momenta:
\begin{align}
 C_N^1=&\frac{1}{3}\left(C_N(2\pi/L,0,0)+C_N(0,2\pi/L,0)+C_N(0,0,2\pi/L)\right),\\
 C_N^2=&\frac{1}{6}\left(C_N(0,2\pi/L,2\pi/L)+C_N(0,-2\pi/L,2\pi/L)+C_N(2\pi/L,0,2\pi/L) \right.\nonumber\\
	&\left. {} + C_N(2\pi/L,0,-2\pi/L)+C_N(2\pi/L,2\pi/L,0)+C_N(2\pi/L,-2\pi/L,0)\right).
\end{align}

\begin{figure}[t]
\subfigure[]{\includegraphics[width=0.47\textwidth,clip]{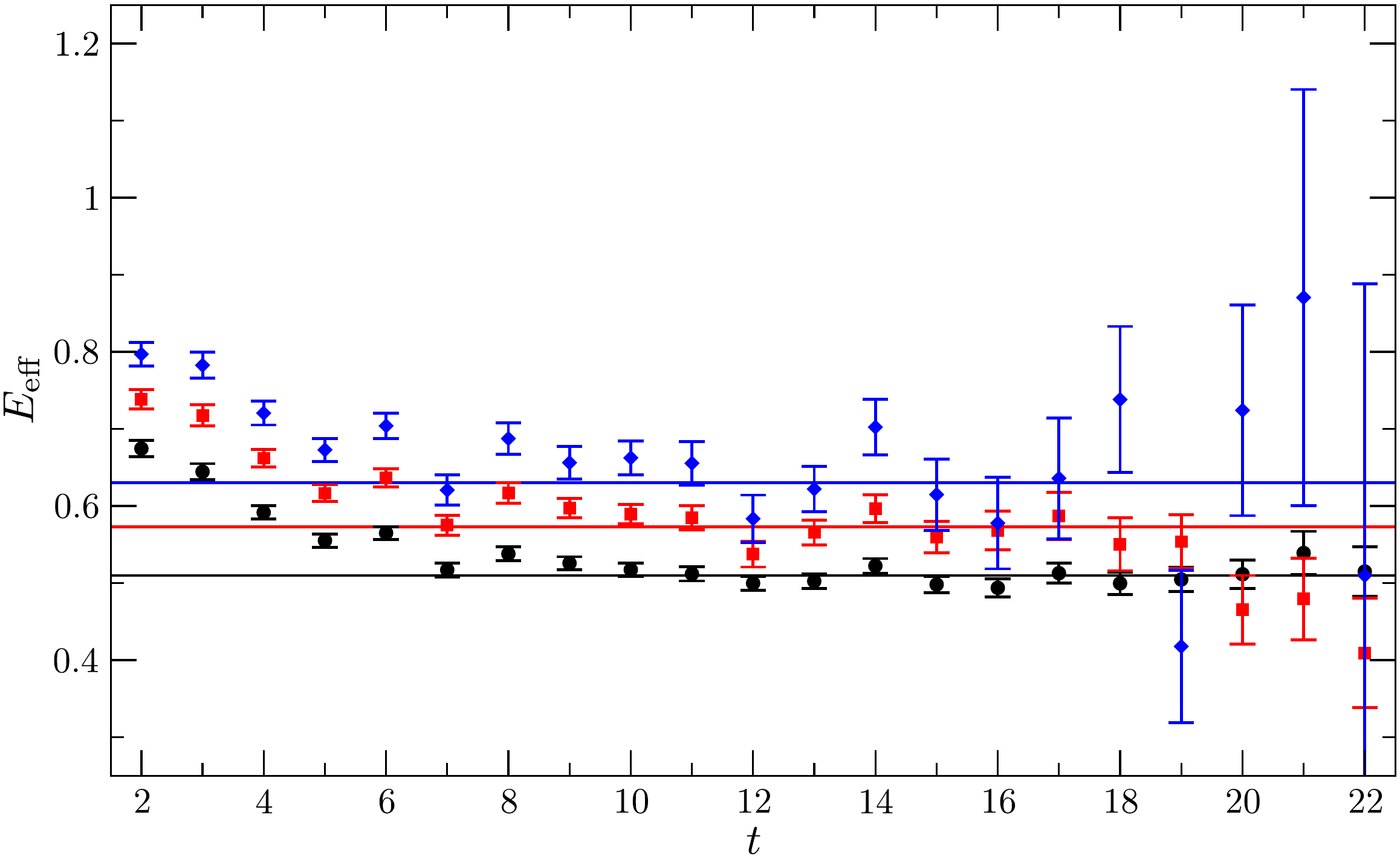}}
\subfigure[]{\includegraphics[width=0.47\textwidth,clip]{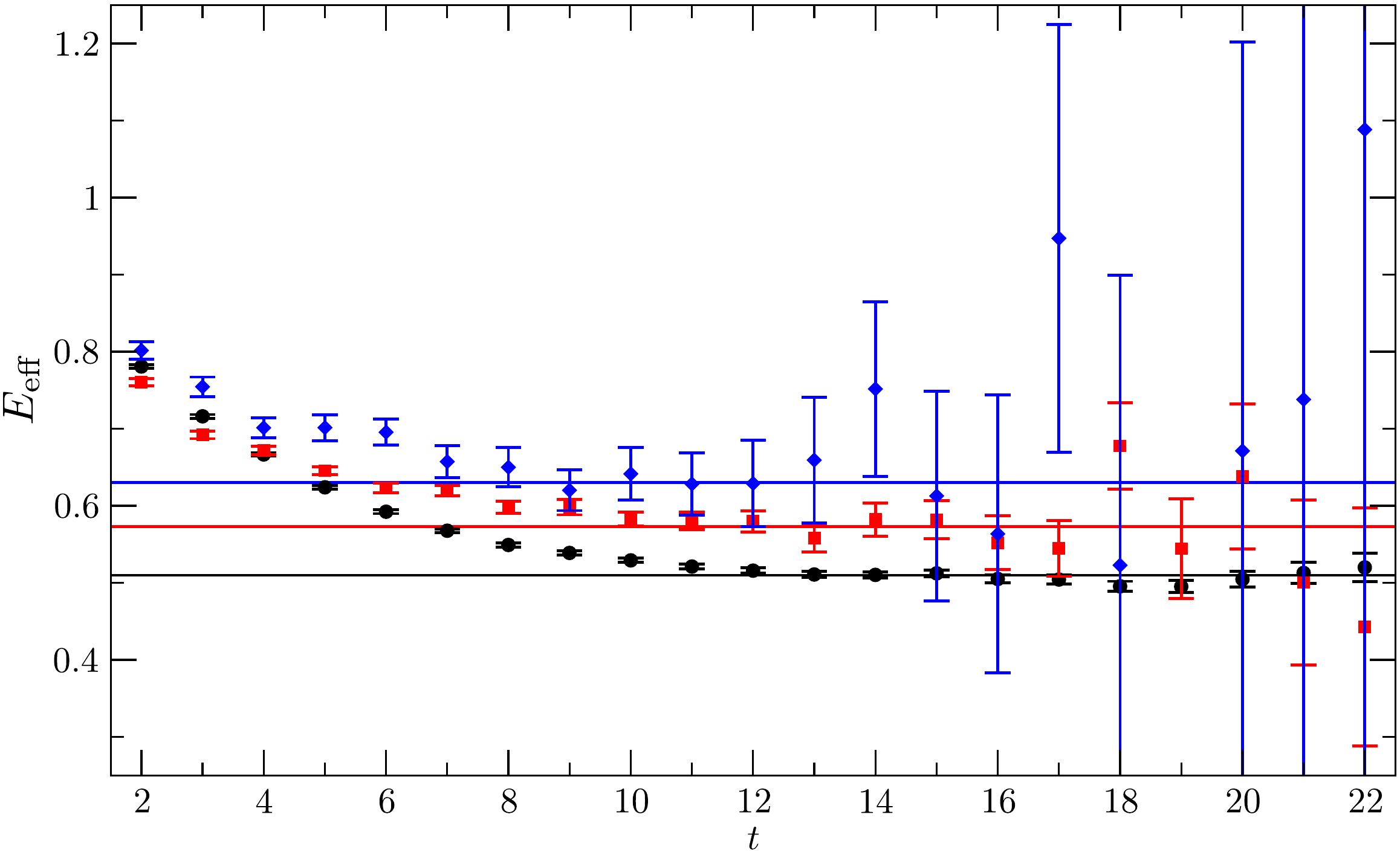}}
\caption{\label{fig:effmass} 
Effective energy plots for different nucleon momenta at $\beta=5.40$ 
and $\kappa=0.13610$ for the nucleon correlator (a) and the 
distribution amplitude correlators (b), where we have averaged over 
all available correlators. The black circles were obtained at 
zero nucleon momentum, the red squares and 
blue diamonds correspond to $\vec p {\,}^2=(2\pi/L)^2$ and 
$\vec p {\,}^2 =2 (2\pi/L)^2$, respectively. The lowest black line shows 
the nucleon mass as obtained by direct calculation. The middle red and the 
top blue line correspond to energies 
$E^2_\mathrm{eff}=m^2_\mathrm{eff}+\vec p {\,}^2$ with 
$\vec p {\,}^2 =(2\pi/L)^2$ and $\vec p {\,}^2 =2 (2\pi/L)^2$, respectively. }
\end{figure}

As already mentioned, the location of the effective mass plateaus 
for the nucleon correlator  differs from that for the other 
correlators as exemplified in
Fig.~\ref{fig:effmass}. Thus, instead of calculating the ratios of the
correlators we have performed a joint fit. As all correlators are evaluated on
the same gauge configuration we should also take into account all possible
statistical correlations. We have employed two different fitting 
procedures with different possibilities for incorporating the correlations:
\begin{itemize}
 \item[\bf PC: ] The first possibility is to fit every moment of the 
DA separately, e.g., for $f_N\phi^{100}$ we fit the
correlators $C^{100}_1$ and $C^1_{N}$ simultaneously and incorporate the
correlations of both correlators and those between different time-slices.
However, since we want to extract $\phi^{100}$ and not $f_N\phi^{100}$ 
we should in principle also consider the correlation with  $C^{000}$. 
Due to the omission of these additional correlations we call this 
procedure ``Partially Correlated''.
 \item[\bf FC: ] For the second possibility we have estimated the full 
crosscorrelation matrix and call this method therefore ``Fully Correlated''. 
In this case we fit simultaneously 
the correlators for the zeroth, first and second moment as well as the nucleon 
correlator with the same modulus of the momentum.
\end{itemize}

\begin{figure}[ht]
\centering
     \includegraphics[width=0.70\textwidth,clip]{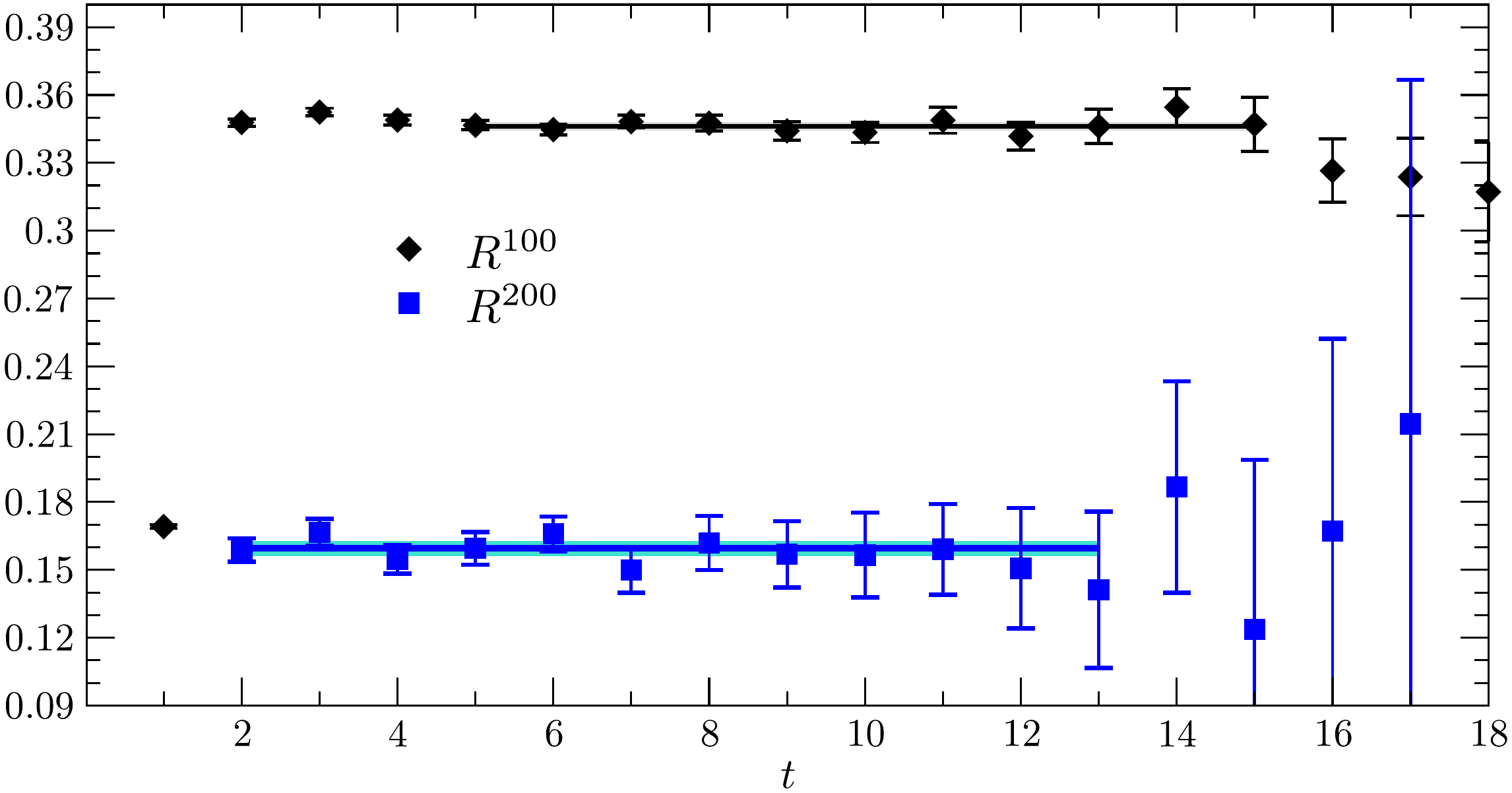}
 \caption{
Plateaus of correlator ratios $R^{100}$ (black diamonds) and $R^{200}$ (blue
squares) for $\beta=5.40$ and $\kappa=0.1361$ together with the corresponding
fit values and the associated error bands.
\label{fig_cratios}
}
\end{figure}

Both methods have some common disadvantages. In order to 
extract the moments we have to perform multiparameter fits which 
involve nucleon mass, different normalization constants and the 
moments. The second disadvantage is the required knowledge of the 
smeared-smeared nucleon correlator for nonzero 
spatial momenta, which introduces additional noise. This requirement can be 
avoided if we consider ratios of the correlators, 
which are equal to ratios of moments:
\begin{align}
 l+&m+n=1: & R^{lmn}=\frac{\phi^{lmn}}{S_1}=\frac{C_1^{lmn}}{C_{S,1}},& &    S_1&=\phi^{100}+\phi^{010}+\phi^{001},\nonumber\\
   &       &                                                          & &C_{S,1}&=C_1^{100}+C_1^{010}+C_1^{001},\label{eq:sum1mom}\\
 l+&m+n=2: 
     &R^{lmn}=\frac{\phi^{lmn}}{S_2}=\frac{C_2^{lmn}}{C_{S,2}},& &S_2    &=2(\phi^{011}+\phi^{101}+\phi^{110})+\phi^{200}+\phi^{020}+\phi^{002},
     \nonumber\\
   & &                                                         & &C_{S,2}&=2 (C_2^{011}+C_2^{101}+C_2^{110})+C_2^{200}+C_2^{020}+C_2^{002}.
     \label{eq:sum2mom}
\end{align}
Now we need additional input to determine the normalization of the moments 
$\phi^{lmn}$ with $l+m+n\geq 1$. This can be obtained by using the 
constraint~\eqref{eq:onesum}. Thus, we require, e.g., for the first 
moments that the renormalized moments satisfy
\begin{equation}
 \sum_{ij} Z_{ij} \phi^\mathrm{lat}_j=1,
\end{equation}
where $\phi^\mathrm{lat}_i$ are the unrenormalized lattice values
\begin{equation}
 \phi^\mathrm{lat}_1:=\phi^{100},\quad \phi^\mathrm{lat}_2:=\phi^{010},\quad
\phi^\mathrm{lat}_3:=\phi^{001}
\end{equation}
and $Z$ is the renormalization matrix.
This leads immediately to a constraint for the ratios
$R_i^\mathrm{lat}=\phi^\mathrm{lat}_i/\sum_j \phi^\mathrm{lat}_j$:
\begin{equation}
 \sum_i \phi^\mathrm{lat}_i=\frac{1}{\sum_{ij} Z_{ij} R^\mathrm{lat}_j}.
\end{equation} 
As in this case we use explicitly the constraint \eqref{eq:onesum} we call 
this analysis method ``constrained''. The calculation of the ratios $R^{lmn}$ 
does not suffer from the disadvantages mentioned above. 
Fitting these ratios to 
a constant we can reach a much higher precision compared to the unconstrained
method discussed before. In Fig.~\ref{fig_cratios} we present some of these
ratios obtained on one of the ensembles with $\beta=5.40$. They exhibit longer
and less noisy plateaus compared to the correlators in Fig.~\ref{fig:effmass}.

The lattice results are obtained at nonphysical quark masses and we have to 
extrapolate them to the physical point. To our knowledge there are no 
calculations in chiral perturbation theory to guide our extrapolation. 
Therefore we have to rely on the behavior of our data and extrapolate them 
linearly to the physical point. To estimate the systematic uncertainty of 
this chiral extrapolation we have performed also an extrapolation 
including a quadratic term. The systematic uncertainty is then taken 
to be the difference of the two results.

In the following we present the results of the constrained and unconstrained 
analysis methods discussed before in the \msb\ scheme at $4\,\mathrm{GeV}^2$ 
while the raw lattice results are summarized in Appendix~\ref{app:latres}. 
Using the unconstrained analysis we obtain the normalization constants of the 
DAs and test how good the constraint in \eqref{eq:sumrule}
is satisfied. Better results with smaller errors for the higher moments of the
leading-twist DA are then obtained from the constrained
analysis.

\subsection{Unconstrained analysis}

In Table~\ref{tab_chi_fc} we present the results for the different constants
which are associated with operators without derivatives: the nucleon wave
function normalization constant $f_N$ and the next-to-leading twist
normalization constants $\lambda_1$ and $\lambda_2$. Our results confirm the
relative signs of $f_N$, $\lambda_1$ and $\lambda_2$ calculated in
\cite{Braun:2000kw,Kolesnichenko:1984dj}. Furthermore we observe
$m_N(2\lambda_1+\lambda_2)/8=\alpha+\beta \approx 0$ as in
\cite{Aoki:2006ib,Aoki:2008ku}. This is expected since  due to
\eqref{eq:lambdasum} $2\lambda_1+\lambda_2$ vanishes in the nonrelativistic
limit and is
known to be small at small quark masses \cite{Sasaki:2001nf}.
\begin{figure}[t]
\subfigure[\label{fig_mom0_fn}]{\includegraphics[width=0.47\textwidth,clip]{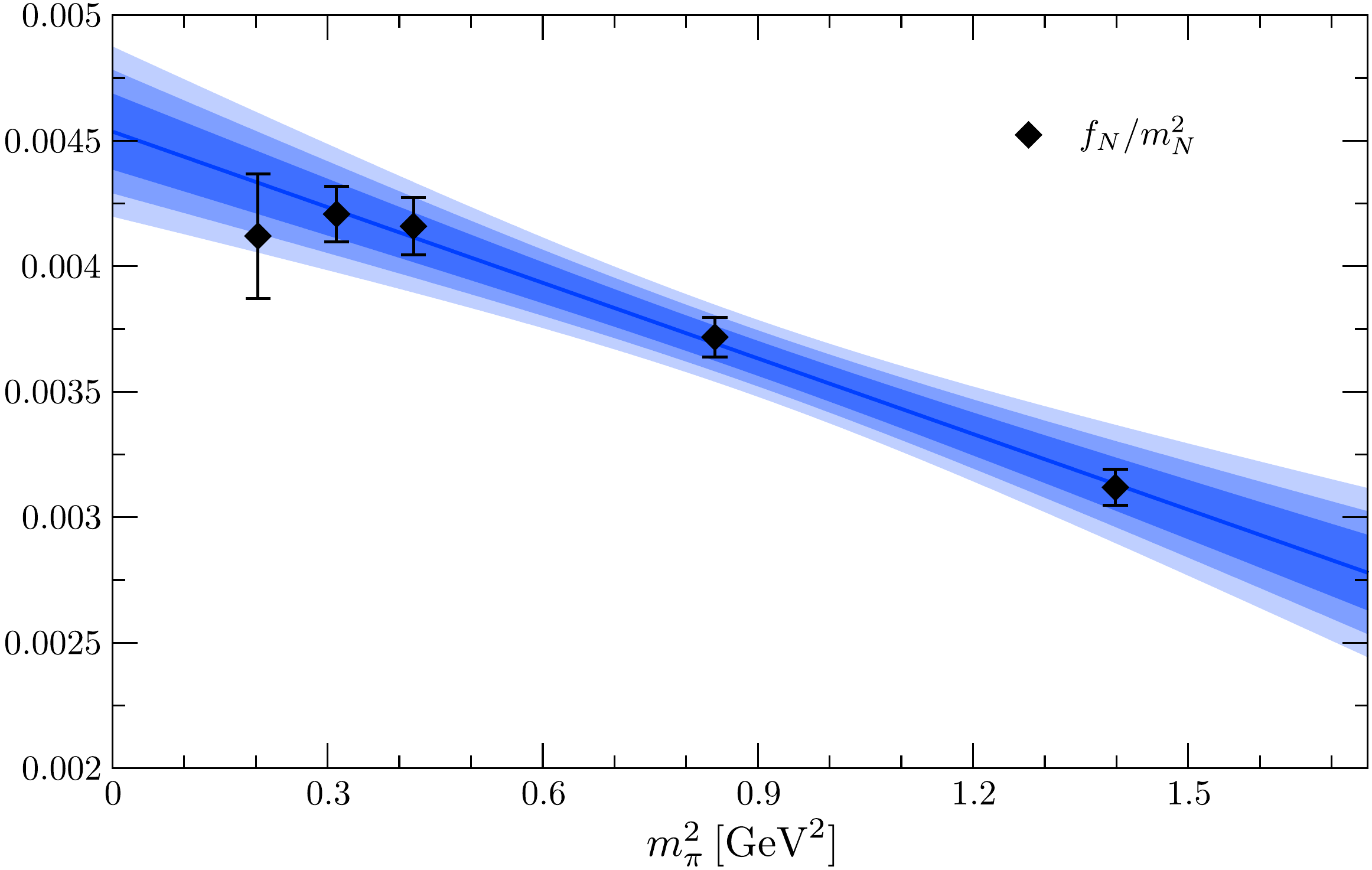}}
\subfigure[\label{fig_mom0_li}]{\includegraphics[width=0.47\textwidth,clip]{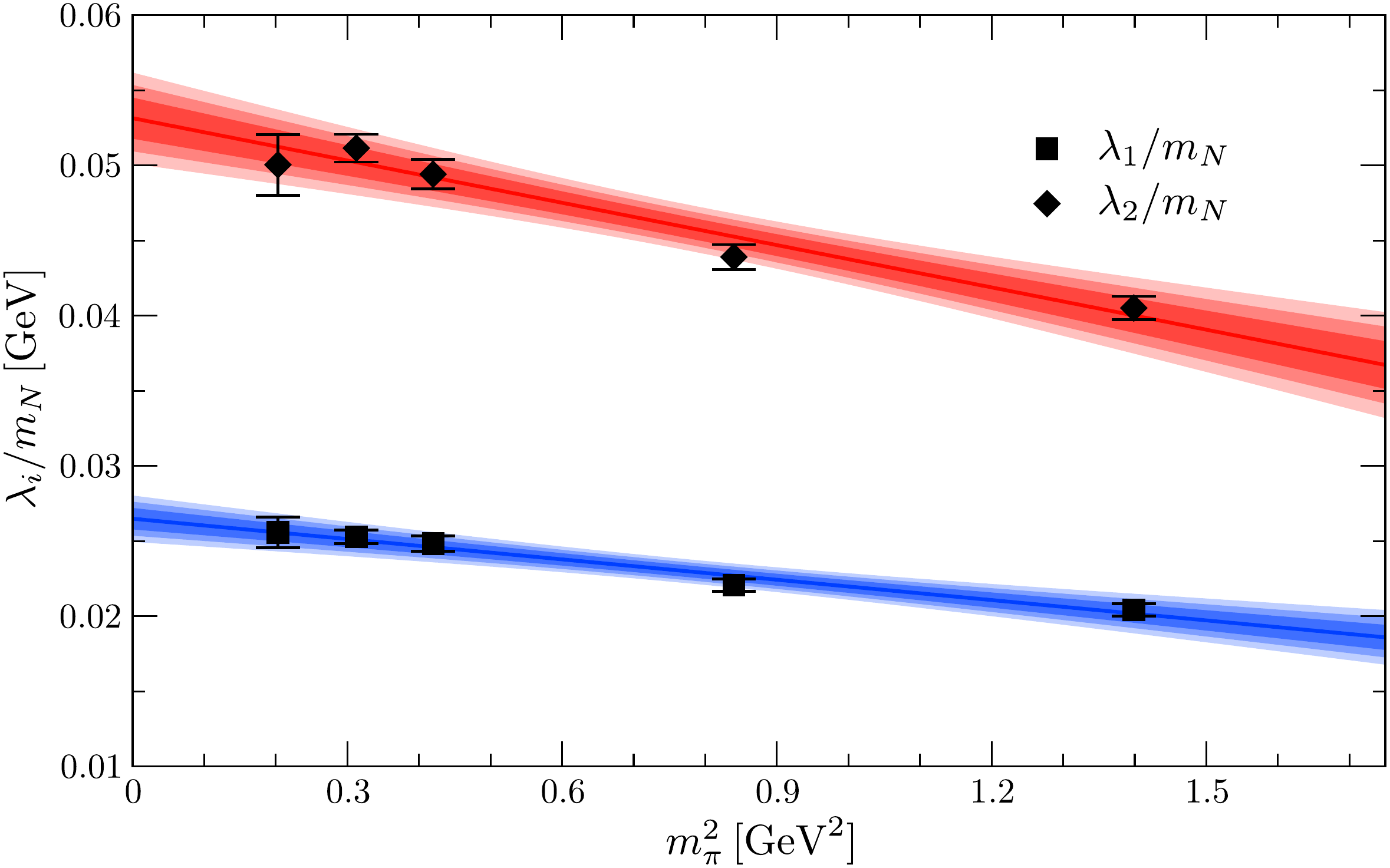}}
\caption{\label{fig_mom0} Linear chiral extrapolation for $f_N/m_N^2$ (a) and
$\lambda_i/m_N$ (b) with the 1,2 and 3 sigma error bands. }
\end{figure}

Our results for the nucleon wave function normalization constant $f_N$ exhibit 
a clearly nonlinear behavior as a function of $m_\pi^2$. However, the
dimensionless ratio $f_N/m_N^2$ is approximately linear  (see
Fig.~\ref{fig_mom0_fn}) and it has the additional advantage that it does not
suffer from the uncertainty in setting the scale on the lattice. The chiral
behavior of $\lambda_1$ and $\lambda_2$ is less clear and we have performed two
different chiral extrapolations for these quantities. 
First we have extrapolated the constants $\lambda_i$ linearly to the
physical point and then we have applied the
same procedure to the ratios $\lambda_i/m_N$. The linear fit looks more
favorable for the ratios $\lambda_i/m_N$ (see Fig.~\ref{fig_mom0_li}). Thus we
take the results from this fit as our final values, but for comparison we also
give the results from the other extrapolation.
In contrast to \cite{Aoki:2006ib,Aoki:2008ku} we do not observe linear
behavior for $m_N \lambda_i$ as a function the quark mass. However, our results
from the linear extrapolation of $\lambda_i/m_N$ are compatible within the
errors with those in \cite{Aoki:2006ib,Aoki:2008ku}.

We have determined the moment combinations
$\varphi^{lmn}=2\phi^{lmn}-\phi^{nml}$ also directly and not from the results
for $\phi^{lmn}$, using the PC fitting procedure. Thus we had also to compute
$f_N$ within this approach. We have also determined $\lambda_i$ using this
analysis method. The results are presented in Table~\ref{tab_chi_pc}. The
correlators for higher moments entering the FC fitting procedure seem to favor
slightly larger nucleon masses, while the PC analysis leads to somewhat higher
values of the normalization constants. We consider the values for the
normalization constants obtained within the PC analysis to be more reliable as
they are not perturbed by the noisier correlators for the higher moments.

\squeezetable
\begin{table}[ht]
\centering
\renewcommand{\arraystretch}{1.25}
   \begin{tabular}{  c D{.}{.}{18}  D{.}{.}{18} }
\hline\hline
 $\beta$                                     & \multicolumn{1}{c}{5.40} & \multicolumn{1}{c}{5.29}   \\
\hline
$ f_N/m_N^2 \cdot 10^{3}$                    &  3.486(60)(56)(60)       & 3.290(62)(100)(72)  \\
$-\lambda_1/m_N \cdot10^{3}[\mathrm{GeV}]$   & 40.64(65)(194)(110)      & 41.24(72)(200)(128)   \\
$-\lambda_1 \cdot 10^{3}[\mathrm{GeV}^2]$    & 49.84(95)(290)(135)      & 52.47(104)(135)(164)   \\
$ \lambda_2/m_N \cdot 10^{3} [\mathrm{GeV}]$ & 80.17(131)(396)(218)     & 82.08(146)(452)(254) \\
$ \lambda_2 \cdot 10^{3} [\mathrm{GeV}^2]$   & 98.53(189)(601)(268)     & 105.12(209)(250)(324) \\
\hline
  $\phi^{100}$                               & 0.3457(75)(89)(3)        & 0.3530(62)(132)(7) \\
  $\phi^{010}$                               & 0.3124(81)(128)(4)       & 0.3176(62)(108)(2) \\
  $\phi^{001}$                               & 0.3142(77)(100)(4)       & 0.3283(62)(68)(4) \\
\hline
  $\phi^{011}$                               & 0.0838(73)(266)(44)      & 0.0851(61)(1)(44)   \\
  $\phi^{101}$                               & 0.1121(92)(250)(58)      & 0.1020(66)(179)(68)   \\
  $\phi^{110}$                               & 0.1051(67)(6)(4)         & 0.0979(54)(5)(9)   \\
  $\phi^{200}$                               & 0.1523(106)(699)(129)    & 0.1639(86)(216)(114)   \\
  $\phi^{020}$                               & 0.1268(97)(153)(98)      & 0.1277(79)(1)(76)   \\
  $\phi^{002}$                               & 0.1398(99)(45)(128)      & 0.1473(84)(40)(111)   \\
\hline
\end{tabular}
\caption{ \label{tab_chi_fc}
Chirally extrapolated results from the FC analysis for normalization constants
and the moments $\phi^{lmn}$ at $\beta=5.40$ and $\beta=5.29$ in the \msb\
renormalization scheme at $4\,\mathrm{GeV}^2$. The first error is the combined
statistical error of the  moments and renormalization matrices. The second
(third) errors are the systematic uncertainties due to the chiral extrapolation
(renormalization).
}
\end{table}

\begin{table}[ht]
\centering
\renewcommand{\arraystretch}{1.25}
\begin{tabular}{ c D{.}{.}{18} D{.}{.}{18}}
\hline\hline
 $\beta$                                       &\multicolumn{1}{c}{5.40}& \multicolumn{1}{c}{5.29}\\
\hline
$ f_N/m_N^2 \cdot 10^{3}$                      & 3.672(78)(90)(63)      & 3.538(79)(283)(77)      \\
$-\lambda_1/m_N \cdot 10^{3} [\mathrm{GeV}]$   & 42.19(81)(86)(115)     & 45.07(92)(315)(140)     \\
$ \lambda_2/m_N \cdot 10^{3} [\mathrm{GeV}]$   & 82.91(171)(18)(225)    & 86.90(87)(641)(261)     \\
\hline
  $\varphi^{100}$                              & 0.3871(313)(528)(4)    & 0.3903(204)(464)(12)    \\
  $\varphi^{010}=\phi^{010}$                   & 0.3150(226)(290)(720)  & 0.3298(159)(118)(608)   \\
  $\varphi^{001}$                              & 0.3155(272)(453)(2)    & 0.3277(190)(270)(5)     \\
\hline
  $\varphi^{011}$                              & 0.0712(180)(127)(92)   & 0.0827(137)(103)(92)    \\
  $\varphi^{101}=\phi^{101}$                   & 0.1091(112)(138)(64)   & 0.1176(105)(171)(64)    \\
  $\varphi^{110}$                              & 0.1266(178)(82)(40)    & 0.1069(137)(103)(49)    \\
  $\varphi^{200}$                              & 0.1879(250)(942)(135)  & 0.1709(184)(569)(121)   \\
  $\varphi^{020}=\phi^{020}$                   & 0.1275(149)(105)(108)  & 0.1261(117)(78)(75)     \\
  $\varphi^{002}$                              & 0.1357(233)(375)(135)  & 0.1249(193)(296)(109)   \\
\hline
\end{tabular}
\caption{\label{tab_chi_pc}
Chirally extrapolated PC results for normalization constants and the 
moments $\varphi^{lmn}$ at $\beta=5.40$ and $\beta=5.29$ in the \msb\
renormalization scheme at $4\,\mathrm{GeV}^2$. The first error is the combined
statistical error of the  moments and renormalization matrices. The second
(third) errors are the systematic uncertainties due to the chiral extrapolation
(renormalization). Note that only the values for $\varphi^{lml}$ 
can be directly compared with the values for $\phi^{lml}$ in 
Table~\ref{tab_chi_fc}.
}
\end{table}

\begin{figure}
     \subfigure[\label{fig:mom1_a}]{
          \includegraphics[clip,width=0.47\textwidth]{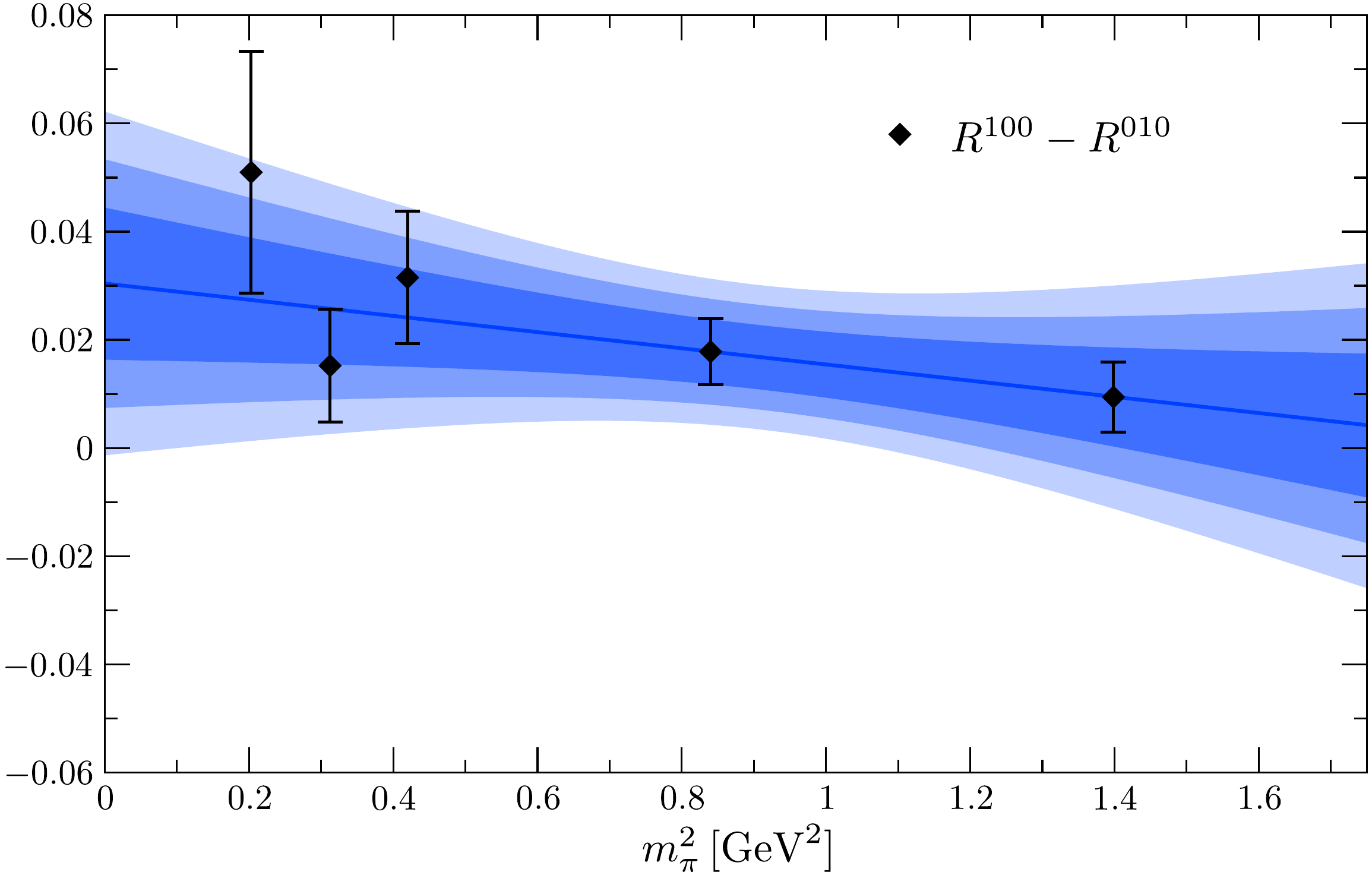}
     }
     \subfigure[\label{fig:mom1_b}]{
          \includegraphics[width=0.47\textwidth,clip]{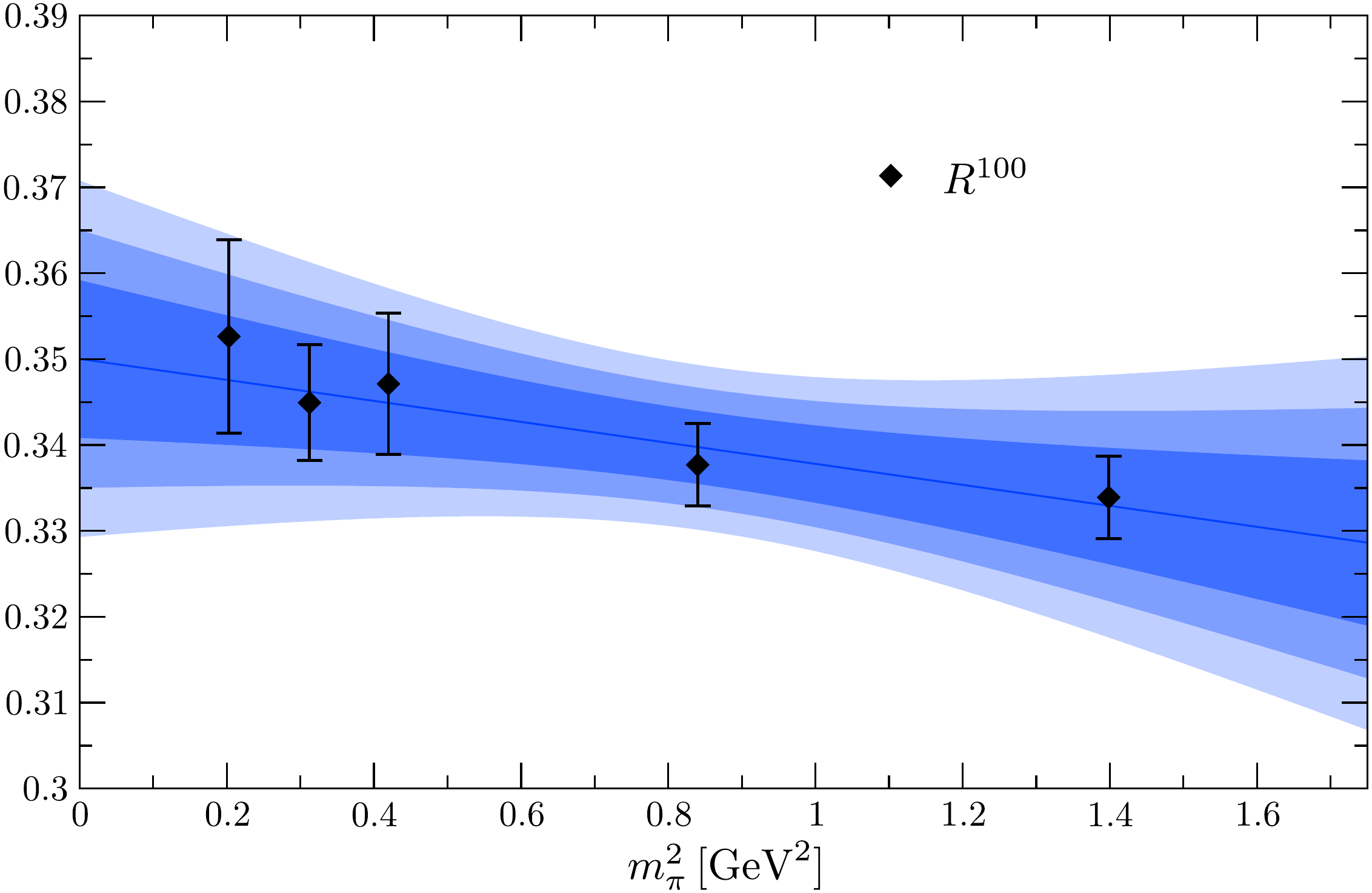}
     }
 \caption{\label{fig:mom1} 
Chiral extrapolation of the asymmetry $(\phi^{100}-\phi^{010})/S_1$ (a) from PC
results and the ratio $\phi^{100}/S_1$ (b) from FC results. We have normalized
the values by $S_1$ so that we are able to compare these directly with the
plots in the constrained analysis in Fig.~\ref{fig:cmom1all}.}
\end{figure}

As expected, the nonzero spatial momenta make the results for the 
first moments noisier than for operators without derivatives.  
The renormalized results for the moments $\phi^{100},\phi^{010}$ 
and $\phi^{001}$ show clearly the deviation
from the asymptotic case  with $\phi^{100}=\phi^{010}=\phi^{001}=1/3$. As the
relative differences of these moments describe the deviation from the symmetric
case, they are of particular interest in phenomenological applications. Thus we
have also determined these differences directly and the bare results 
from the PC analysis are given in Appendix~\ref{app:latres}. 
Although these results show a
significant deviation from the symmetric case, the errors are large and do not
allow reasonable quantitative conclusions. To illustrate these we show in
Fig.~\ref{fig:mom1_a} the most important asymmetry $\phi^{100}-\phi^{010}$
normalized by the sum $S_1$ so that we can compare this later directly with the
results from the constrained analysis. However, the results for the moments are
less affected by the noise as shown on the example of $\phi^{100}$ in
Fig.~\ref{fig:mom1_b} also normalized by $S_1$. 

\begin{figure}
       \subfigure[]{
              \includegraphics[clip,width=0.47\textwidth]{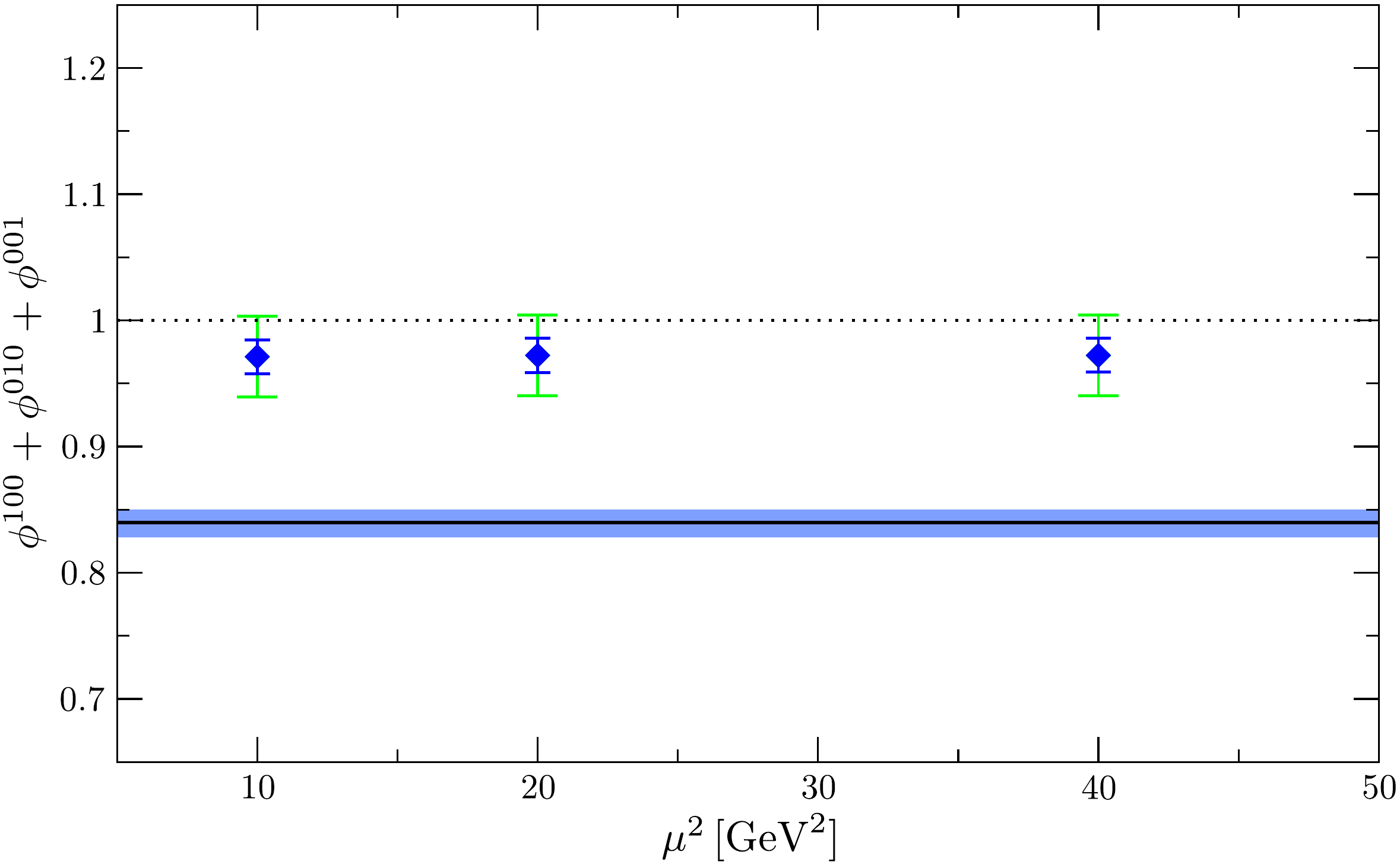}
       }
       \subfigure[]{
              \includegraphics[clip,width=0.47\textwidth]{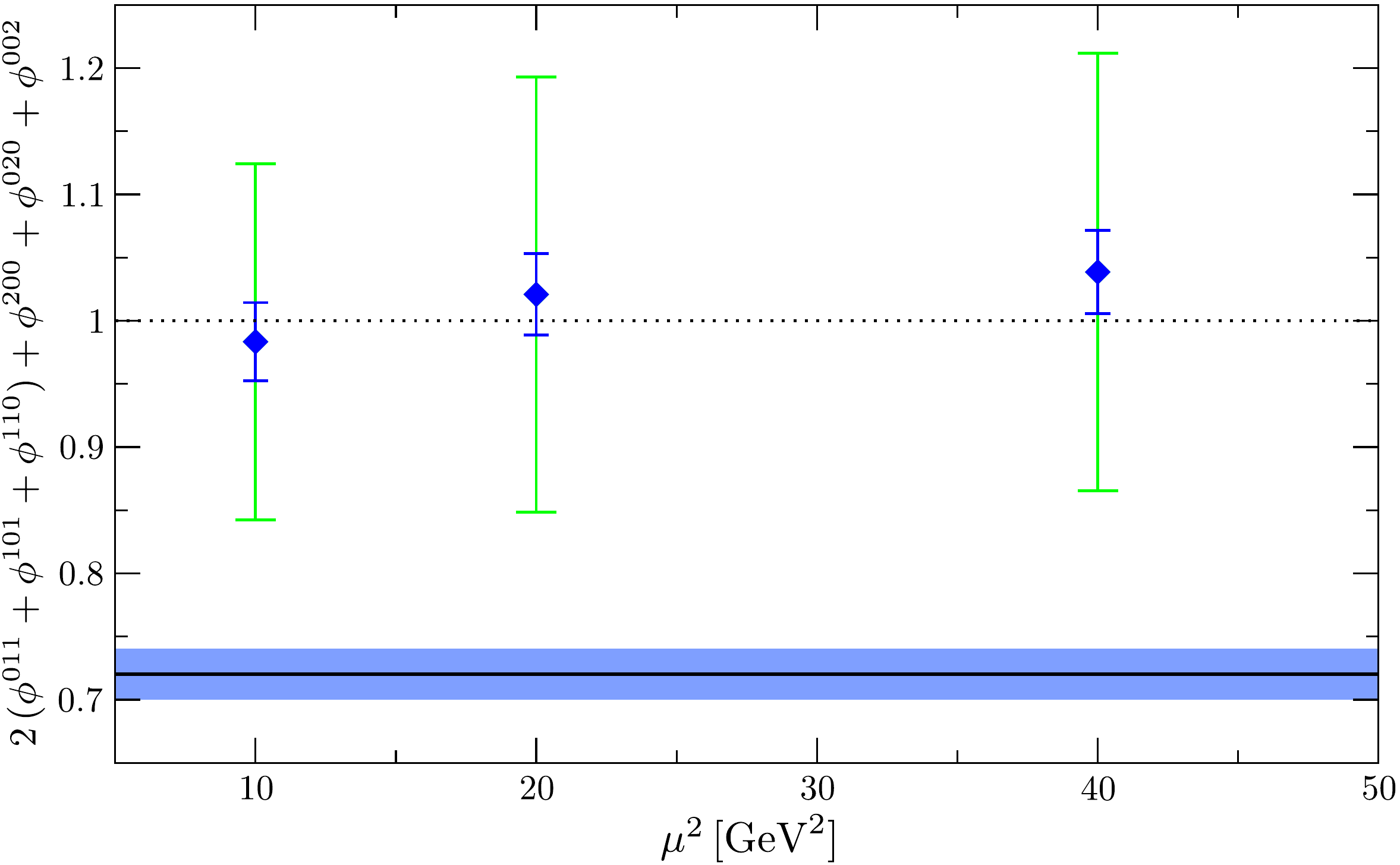}
       }
 \caption{\label{fig:srconstraint} 
The bare (solid black line with statistical error band) and renormalized
(blue diamonds)  sum  of the first moments (a) and second moments (b) according
to Eq.~\eqref{eq:sumrule} as obtained from the FC analysis. The smaller errors
for the renormalized values are purely statistical, while the larger are ones
include the systematical error due to the chiral extrapolation. The three
different points were obtained from three different renormalization 
scales $\mu$ in the RI$^\prime$-MOM scheme to estimate the systematic 
uncertainty due to the renormalization. The theoretical constraint 
\eqref{eq:onesum} that the sum should be exactly equal to one is 
fulfilled in both cases.
}
\end{figure}

We have checked our results by calculating the sums $S_1$ and $S_2$ according 
to Eqs.~\eqref{eq:sum1mom} and \eqref{eq:sum2mom}. The results for the bare and
renormalized sums are shown Fig.~\ref{fig:srconstraint}. For the renormalized
moments the constraint \eqref{eq:sumrule} is fulfilled very well indicating the
consistency of our results. Of course the statistical and systematic errors for
the case of two derivatives in the operators are higher. Nevertheless, the
results still allow us to see the asymmetries. Because of the large errors we
give these only for the bare results in Appendix~\ref{app:latres}.

\subsection{Constrained analysis of higher moments}

In the last section we have seen that the unconstrained analysis of our data 
gives us results consistent with theoretical constraint \eqref{eq:onesum}. 
However, better estimates of moments and in particular of asymmetries 
can be obtained from  the correlator ratios $R^{lmn}$. 
Indeed, the values extracted from the ratios (summarized in 
Table~\ref{tab_chi_cf}) have smaller errors than those from 
the unconstrained analysis. The main reasons for this improvement 
are that we do not have to determine the energy $E(\vec p)$ and
normalization constant $Z_N(\vec p)$ for nonzero spatial momenta as both drop
out in the constrained analysis. This reduces also the statistical
noise as the nucleon correlator with smeared source and sink is not involved
anymore in the data analysis. 

The normalization constants $f_N$ and $\lambda_i$ in
Table~\ref{tab_chi_cf}  were determined by performing a 
joint fit of all relevant correlators. This approach is equivalent to the
FC analysis method. However, as the correlators with higher momenta are
not involved the obtained results have smaller errors compared to the FC
analaysis.  Our values for $\alpha=-0.0091\pm 0.0002_\mathrm{st} \pm
0.0003_\mathrm{sys}$ and $\beta=0.0090\pm 0.0002_\mathrm{st} \pm
0.0003_\mathrm{sys}$ obtained from $\lambda_i/m_N$ at $\beta=5.40$ (see 
Table~\ref{tab_chi_cf}) are consistent within the errors with the recent
results $\alpha=-0.0112\pm 0.0012_\mathrm{st} \pm 0.0022_\mathrm{sys}$ and
$\beta=0.00120\pm 0.0013_\mathrm{st} \pm 0.0023_\mathrm{sys}$ from simulations
with $2+1$ flavors of domain-wall fermions \cite{Aoki:2008ku}.

\begin{figure}[ht]
 \subfigure[]
     {\label{fig:cmom100}
     \includegraphics[width=0.47\textwidth,clip]{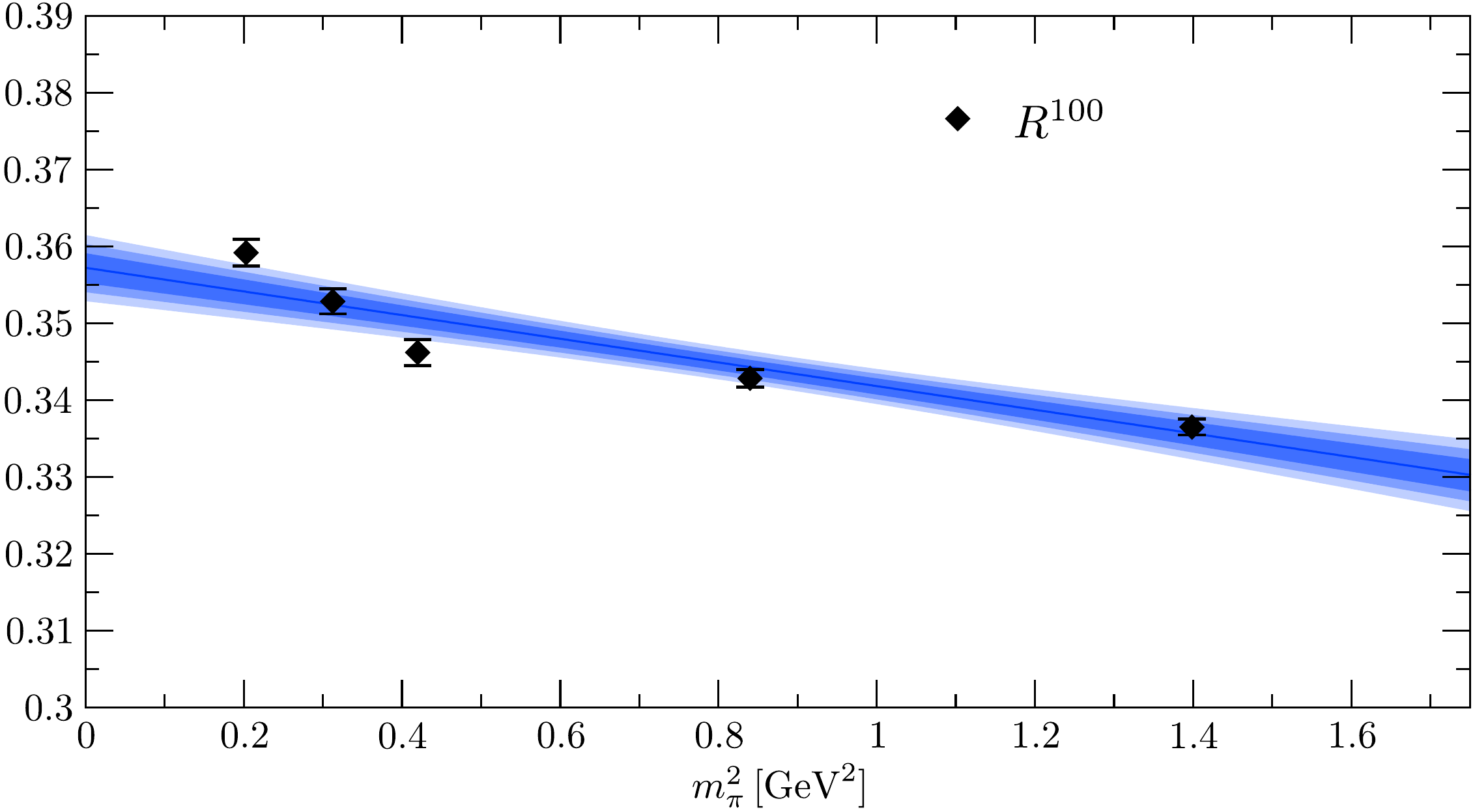}
     }
 \subfigure[]
     {\label{fig:cmom100_2}
      \includegraphics[width=0.47\textwidth,clip]{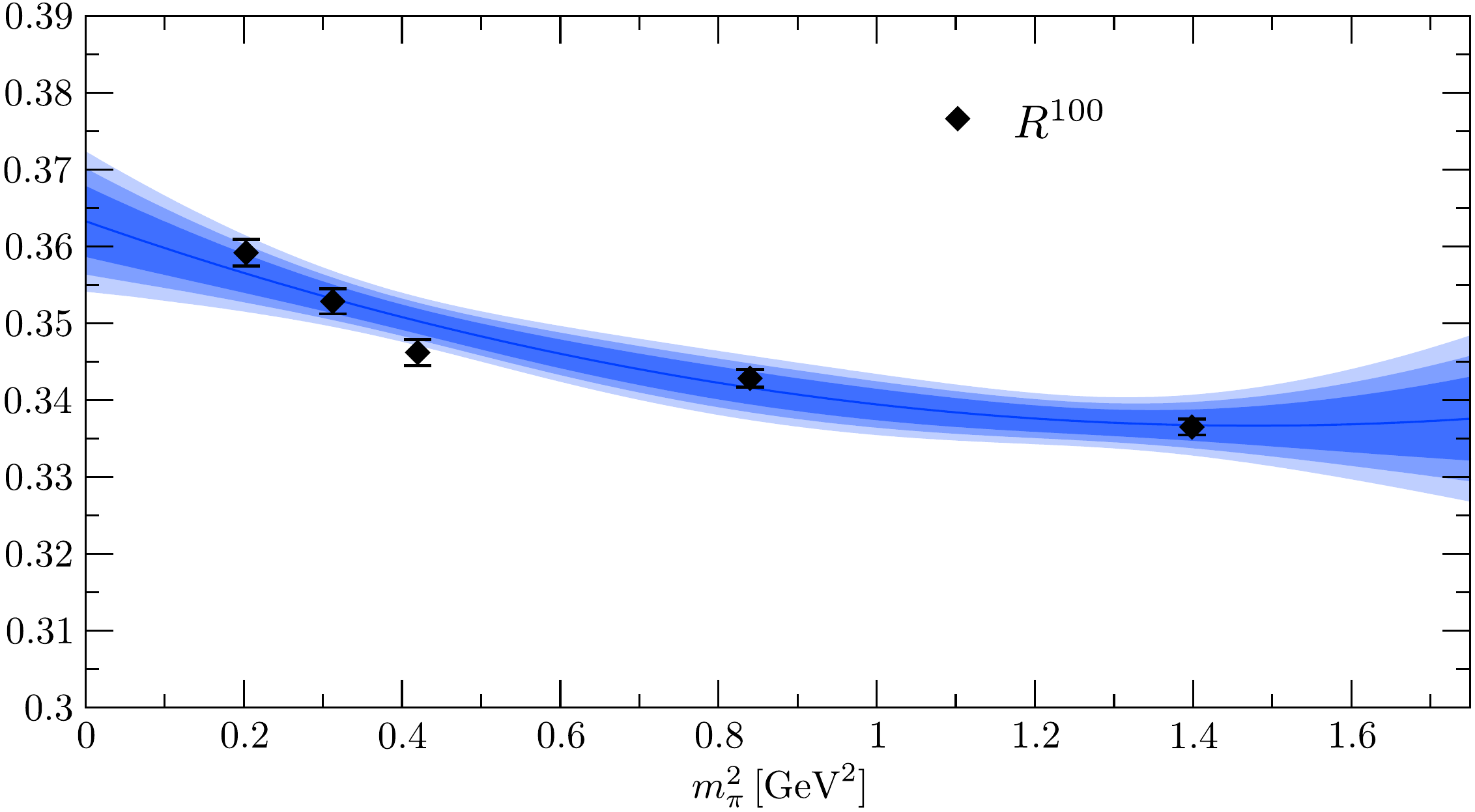}
     }\\
 \subfigure[]
     {\label{fig:cmom1diff}
     \includegraphics[width=0.47\textwidth,clip]{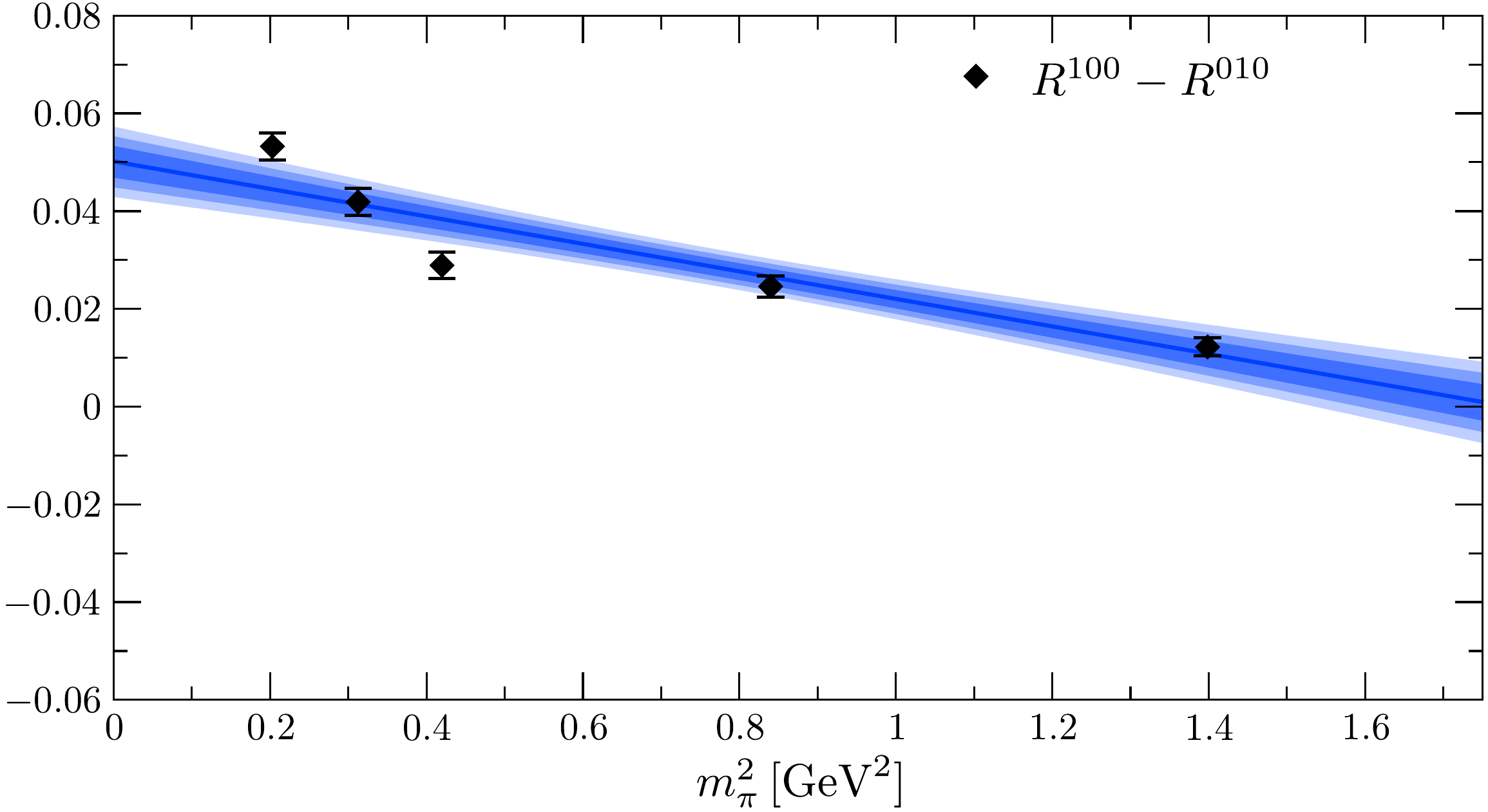}
     }
 \subfigure[]
     {\label{fig:cmom1sum}
      \includegraphics[width=0.47\textwidth,clip]{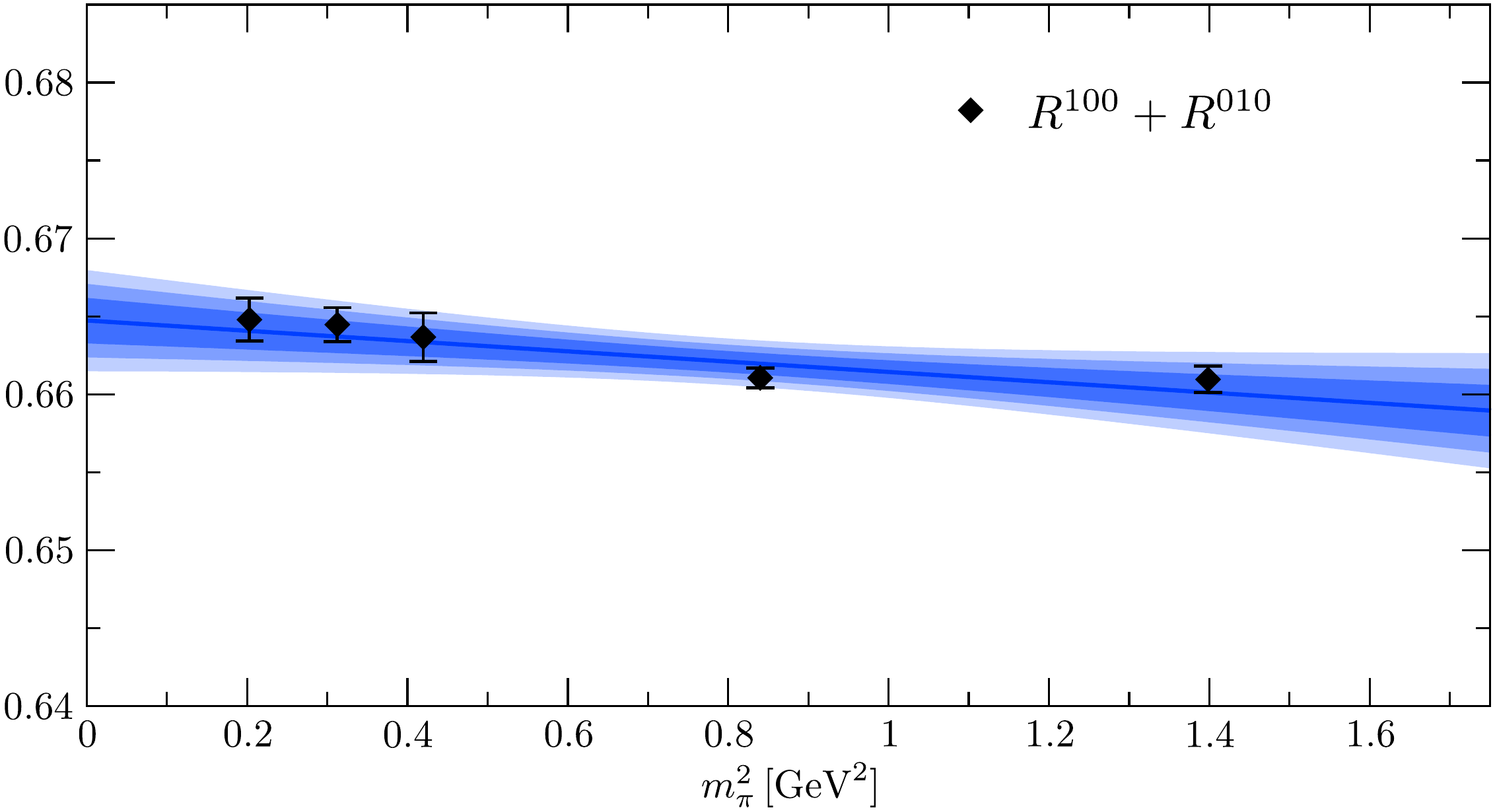}
     }
 \caption[]{\label{fig:cmom1all}
The effect of different chiral extrapolations is demonstrated in the 
case of $R^{100}$ where in (a) a linear fit is performed and in (b) 
a quadratic one. In the lower plots we show the chiral extrapolation 
of the asymmetry $R^{100}-R^{010}$ (c)  and the sum $R^{100}+R^{010}$ (d).
All the plots contain also one, two and three sigma error bands 
of the corresponding fits.
}
\end{figure}
\begin{figure}[ht]
 \subfigure[]
     {\label{fig:cmom110}
      \includegraphics[width=0.47\textwidth,clip]{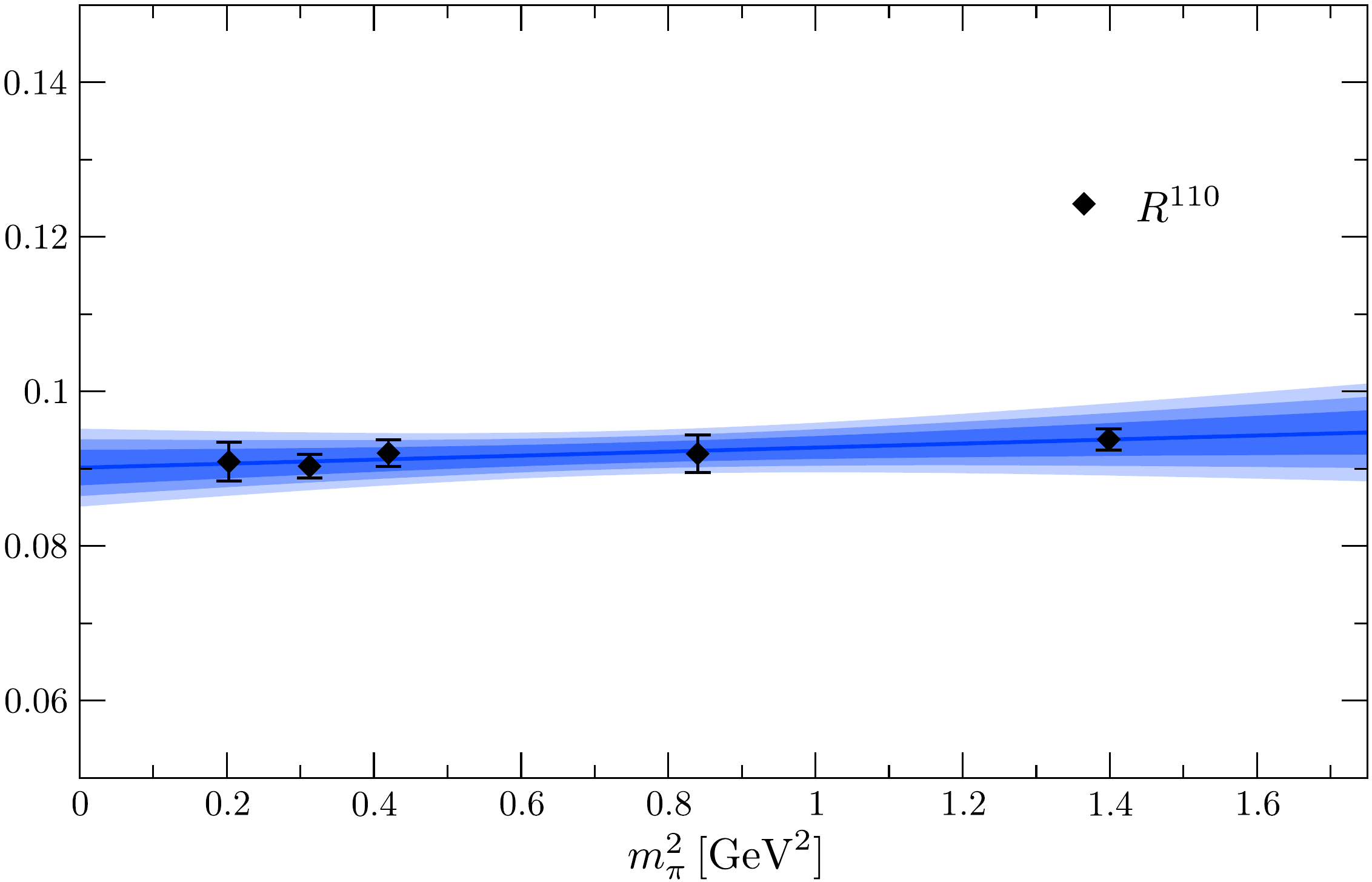}
     }
\subfigure[]
     {\label{fig:mom200mom020}
      \includegraphics[width=0.47\textwidth,clip]{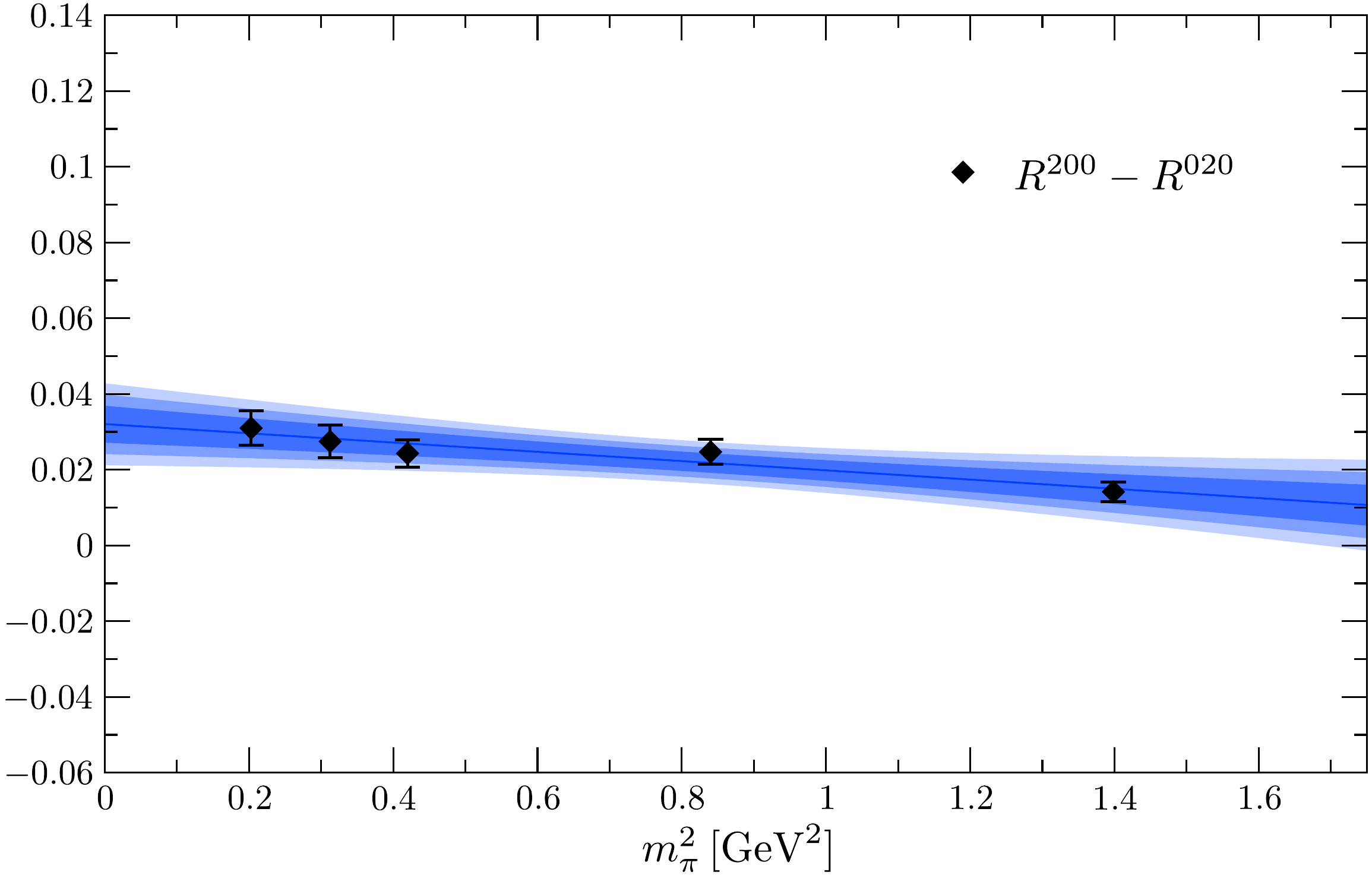}
     }
 \caption[]{\label{fig:cmom2all}
Linear chiral extrapolation of the second moment ratio $R^{110}$ (a) and of the
asymmetry $R^{200}-R^{020}$ (b) as obtained from the constrained analysis with
one, two and three sigma error bands of the corresponding fits.}
\end{figure}

In principle one can calculate similar ratios for correlators involving 
$$\varphi^{lmn}=V^{lmn}-A^{lmn}$$ instead of using 
$$\phi^{lmn}=(V^{lmn}-A^{lmn}+2T^{lnm})/3.$$ 
However, this leads to statistical errors which are about three times larger.

To illustrate the dependence of $R^{100}$ on the pion mass we present in
Figs.~\ref{fig:cmom100} and \ref{fig:cmom100_2} linear and quadratic chiral
extrapolations of this quantity. As $R^{010}$ exhibits a similar behavior, but
with opposite slope, the deviation from a linear dependence is amplified in the
asymmetry $R^{100}-R^{010}$ (Fig.~\ref{fig:cmom1diff}). On the other hand, this
leads to linear behavior of $R^{100}+R^{010}$ (Fig.~\ref{fig:cmom1sum}). Thus,
due to momentum conservation one expects also linear behavior for $R^{001}$,
which is indeed observed in our data. Of course deviations from linear behavior
are also possible for all other moments. However, they seem to be smaller than
present statistical errors. Comparing the chiral extrapolations in
Fig.~\ref{fig:mom1_a} to Fig.~\ref{fig:cmom1diff} and in Fig.~\ref{fig:mom1_b}
to Fig.~\ref{fig:cmom100} reveals the increased accuracy of the constrained
analysis. 

This increase of accuracy is even more important for higher moments. From
Figs.~\ref{fig:cmom110} and \ref{fig:mom200mom020} it is obvious that the
improvement for the second moments allows us not only to determine the moments
but also the more interesting asymmetries. Even more, with the help of the
constraints \eqref{eq:sumrule} the moments $\phi^{200}$, $\phi^{020}$,
$\phi^{002}$ can be calculated from the other second moments and the first
moments. Our results are fully consistent with the direct determination. This
approach can be particularly advantageous in the calculation of the third
moments as one can then dispense with the evaluation of $\phi^{300}$,
$\phi^{030}$, $\phi^{003}$.

\begin{table}[ht]
\renewcommand{\arraystretch}{1.25}
\centering
\begin{tabular}{ c D{.}{.}{18} D{.}{.}{18}}
\hline\hline
 $\beta$                              &\multicolumn{1}{c}{5.40} & \multicolumn{1}{c}{5.29} \\
\hline
$ f_N/m_N^2 \cdot 10^{3}$                   & 3.573(69)(33)(61)     & 3.392(68)(178)(74) \\
$-\lambda_1/m_N \cdot10^{3}[\mathrm{GeV}]$  & 41.29(74)(45)(113)    & 42.32(81)(277)(133) \\
$ \lambda_2/m_N \cdot10^{3}[\mathrm{GeV}]$  & 81.27(149)(90)(221)   & 83.90(167)(599)(261) \\
\hline
  $\phi^{100}$                              & 0.3638(11)(68)(3)     & 0.3549(11)(61)(2)    \\
  $\phi^{010}=\varphi^{010}$                & 0.3023(10)(42)(5)     & 0.3100(10)(73)(1)    \\
  ${\phi^{001}}^\star$                      & 0.3339(9)(26)(2)      & 0.3351(9)(11)(2)     \\
  $\phi^{100}-\phi^{001}$                   & 0.0300(23)(93)(1)     & 0.0199(23)(46)(4)    \\
  $\phi^{001}-\phi^{010}$                   & 0.0313(17)(12)(7)     & 0.0251(16)(84)(2)    \\
\hline
  $\phi^{011}$                              & 0.0724(18)(82)(70)    & 0.0863(23)(97)(74)   \\
  $\phi^{101}=\varphi^{101}$                & 0.1136(17)(32)(21)    & 0.1135(23)(3)(33)    \\
  ${\phi^{110}}^\star$                      & 0.0937(16)(3)(38)     & 0.0953(21)(58)(31)   \\
  $\phi^{200}$                              & 0.1629(28)(7)(68)     & 0.1508(38)(213)(64)  \\
  ${\phi^{020}}^\star=\varphi^{020}$        & 0.1289(27)(37)(51)    & 0.1207(32)(43)(56)   \\
  $\phi^{002}$                              & 0.1488(32)(77)(73)    & 0.1385(36)(47)(64)   \\
  $\phi^{110}-\phi^{011}$                   & 0.0211(27)(78)(32)    & 0.0075(33)(69)(44)   \\
  $\phi^{101}-\phi^{110}$                   & 0.0204(21)(134)(50)   & 0.0172(29)(82)(57)   \\
  $\phi^{200}-\phi^{020}$                   & 0.0321(33)(69)(55)    & 0.0335(43)(26)(78)   \\
  $\phi^{002}-\phi^{020}$                   & 0.0193(24)(32)(42)    & 0.0170(36)(8)(56)    \\
\hline
\end{tabular}
\caption{ \label{tab_chi_cf}
The results for $\phi^{lmn}$ and the relevant asymmetries as obtained 
from the chirally extrapolated ratios $R^{lmn}$  in the \msb\ 
renormalization scheme at $4\,\mathrm{GeV}^2$. The values marked by a 
star were used in the analysis of the corresponding asymmetries to 
determine the overall normalization. The first error is the combined 
statistical error of the  moments and renormalization matrices 
dominated by the statistical uncertainties of the moments. The second 
(third) errors are the systematic uncertainties due to the chiral 
extrapolation (renormalization).
}
\end{table}

Our data do not allow us to perform a continuum extrapolation. However, 
the fact that the $\beta=5.29$ and $\beta=5.40$ results are compatible 
with each other indicates that its effect would be small. Thus we take 
the data from our finer lattice ($\beta=5.40$) as our final numbers. 
For convenience we summarize in Table~\ref{tab:philmn} the corresponding  
moments $\varphi^{lmn}$ at two different renormalization scales as 
obtained from the $\beta=5.40$ results in Table~\ref{tab_chi_cf}. 
The change of scales has been performed in the one-loop approximation
with $\Lambda_{\overline{\mathrm{MS}}} = 226 \, \mbox{MeV}$.
For this purpose the moments $\phi^{lmn}$, being not multiplicatively 
renormalizable, had to be expressed as linear combinations of quantities
that are multiplicatively renormalizable, at least in the one-loop 
approximation, i.e., the coefficients $c_{nl}$ to be introduced in the
next section. Their values (and hence also the values of the moments 
$\phi^{lmn}$ at the new scale) depend somewhat on the set of moments 
$\phi^{lmn}$ that are used as an input. We employed here the set 1 of moments
defined in the following section.

\begin{table}[ht]
\centering 
\renewcommand{\arraystretch}{1.25}
\begin{tabular}{   c  D{.}{.}{7}   D{.}{.}{19}  D{.}{.}{19} }
\hline\hline
                    & \multicolumn{1}{c}{Asymptotic}  & \multicolumn{1}{c}{$\mu^2=4\;\mathrm{GeV}^2$}  & \multicolumn{1}{c}{$\mu^2=1\;\mathrm{GeV}^2$}\\ 
\hline
$f_N\cdot 10^3 [\mathrm{GeV}^2]$&  \multicolumn{1}{c}{$-$}	  & 
3.144(61)(83) & 3.234(63)(86)  \\
$-\lambda_1\cdot 10^3 [\mathrm{GeV}^2]$&\multicolumn{1}{c}{$-$}&  38.72(76)(148)
  & 35.57(65)(136)  \\
$ \lambda_2\cdot 10^3 [\mathrm{GeV}^2]$&\multicolumn{1}{c}{$-$}& 
76.23(139)(291)  & 70.02(128)(268) \\
\hline
  $\varphi^{100}$ & \qquad\frac13\approx 0.333   & \qquad 0.3936(34)(126)       & \qquad 0.3999(37)(139)\\
  $\varphi^{010}$ & \frac13\approx 0.333         &  0.3023(10)(47)              & 0.2986(11)(52)  	\\
  $\varphi^{001}$ & \frac13\approx 0.333         &  0.3041(29)(96)              & 0.3015(32)(106)	\\
\hline
  $\varphi^{200}$ & \frac17\approx 0.143         &  0.1788(53)(179)             & 0.1816(64)(212)	\\
  $\varphi^{020}$ & \frac17\approx 0.143         &  0.1289(27)(88)              & 0.1281(32)(106)	\\
  $\varphi^{002}$ & \frac17\approx 0.143         &  0.1310(95)(324)             & 0.1311(113)(382)	\\
  $\varphi^{011}$ & \frac{2}{21}\approx 0.095    &  0.0659(74)(266)             & 0.0613(89)(319)	\\
  $\varphi^{101}$ & \frac{2}{21}\approx 0.095    &  0.1072(35)(128)             & 0.1091(41)(152)	\\
  $\varphi^{110}$ & \frac{2}{21}\approx 0.095    &  0.1076(56)(182)             & 0.1092(67)(219)\\
\hline
\end{tabular}
\caption{ \label{tab:philmn}
Moments $\varphi^{lmn}$ as obtained from the independent subset $\phi^{010}$,
$\phi^{001}$, $\phi^{110}$, $\phi^{200}$ and $\phi^{020}$ at $\beta=5.40$ in
Table~\ref{tab_chi_cf} at two different scales $\mu^2=4\;\mathrm{GeV}^2$ and 
$\mu^2=1\;\mathrm{GeV}^2$ in the \msb\ renormalization scheme.
}
\end{table}

\section{Modelling the nucleon distribution amplitude \label{sec:modelda}}

Since the available nonperturbative information on the nucleon 
DA comes in the form of a few first moments, it is 
tempting to choose a model which is polynomial in momentum fractions 
at the reference scale $\mu_0$. A natural choice corresponds to the 
(truncated) expansion in contributions of multiplicatively renormalizable
(to leading order) operators of increasing 
dimension~\cite{Braun:1999te,Stefanis:1999wy}:   
\begin{equation}
 \varphi(x_i,\mu^2)=120 x_1 x_2 x_3 \sum_{n=0}^{N} \sum_{l=0}^n c_{nl} P_{nl}(x_i) 
     \left(\frac{\alpha_s(\mu)}{\alpha_s(\mu_0)}\right)^{\gamma_{nl}/\beta_0}.
     \label{eq_daexpansion}
\end{equation}  
Here the first subscript, $n=0,\ldots,N$, is the total number of covariant 
derivatives in the corresponding operator and simultaneously the 
order of the polynomial $P_{nl}(x_i)$. The second subscript, 
$l=0,\ldots,n$, enumerates independent
local operators of the same dimension $D=n+3$.
In this way the scale dependence becomes particularly 
simple and the functional form is preserved under renormalization 
in one-loop accuracy.
In addition, thanks to the conformal symmetry of QCD Lagrangian, the
polynomials $P_{nl}(x_i)$ are mutually orthogonal with respect to the $SL(2,{\mathbb R})$ 
scalar product  
\begin{equation}
 \int [dx]\, x_1 x_2 x_3 \, P_{mk}(x_i) \, P_{nl}(x_i) \propto \delta_{mn}\delta_{kl}\,. 
\label{measure}
\end{equation}
By this reason, the set of moments $\phi^{lmn}$, $l+m+n \le 2$, 
calculated in this work is sufficient to determine uniquely  
all coefficients in (\ref{eq_daexpansion}) up to $N=2$, i.e., 
to second order in the quark momentum fractions. Contributions of 
higher order polynomials correspond to higher dimension operators 
and can be added when the corresponding information becomes available.

In the literature there seems to be no standard convention for the 
normalization of the polynomials $P_{nl}(x_i)$ so we choose the 
simplest expressions (cf.~\cite{Braun:1999te,Stefanis:1999wy}):     
\begin{align}
\varphi (x_1,x_2,x_3,\mu^2) = & 120 x_1 x_2 x_3 
 \Big\{  1     + c_{10} (\mu_0) (x_1 \! - 2 x_2 +x_3) L^{\frac{8}{3\beta_0}}
\nonumber
\\ 
&  
+ c_{11} (\mu_0) (x_1  - x_3) L^{\frac{20}{9\beta_0}}
 + c_{20} (\mu_0) 
\left[ 1 + 7 (x_2 - 2 x_1 x_3 - 2 x_2^2) \right] L^{\frac{14}{3\beta_0}}
\nonumber
\\ 
& + c_{21}  (\mu_0)
\left( 1 - 4 x_2 \right) \left( x_1 - x_3 \right) L^{\frac{40}{9\beta_0}}
 \left. + c_{22}  (\mu_0)
\left[ 3 - 9 x_2 + 8 x_2^2 - 12 x_1 x_3 \right] L^{\frac{32}{9 \beta_0}}
\right\}\,,
\label{eq:model}
\end{align}
where
\begin{equation}
 L\equiv\frac{\alpha_s(\mu)}{\alpha_s(\mu_0)}\,,  \qquad 
\beta_0 = 11 - \frac{2}{3} n_F.
\end{equation}
The scale dependence of the normalization constant is to this accuracy
\begin{equation}
 f_N (\mu)=f_N (\mu_0) L^{\frac{2}{3\beta_0}}.
\end{equation} 

The coefficients $c_{nl}$, $l\leq n$, 
are given in terms of the moments $\phi^{lmn}$ as
\begin{align}
c_{10} & = \frac{7}{2} \left( 3 \left(\phi^{100}+ \phi^{001}\right) -2
\right)\,, 
\\
c_{11} & = \frac{63}{2} \left( \phi^{100} -  \phi^{001} \right) \,, 
\\
c_{20} & =-\frac{126}{5} \left(\phi^{200}+ \phi^{002} + 3 \phi^{101}\right)
           + \frac{18}{5} \left(4+c_{10} \right)\,, 
\\
c_{21} & = 378  \left( \phi^{200}-\phi^{002} \right)  -9 c_{11}\,, 
\\
c_{22} & = \frac{126}{5}  \left( 2 \phi^{200}+ 2 \phi^{002}+ \phi^{101}\right)
          - \frac{21}{5} \left(4+c_{10}\right)\,.
\end{align}
Note that for $N=2$ there are five independent coefficients $c_{nl}$,
which is also the number of independent moments
$\phi^{lmn}$ for $l+m+n\leq2$ due to the constraints \eqref{eq:sumrule}.
In the above expressions we have chosen 
$\phi^{100}$, $\phi^{001}$, $\phi^{101}$, $\phi^{200}$ and $\phi^{002}$ 
to be the independent subset.

\begin{figure}[ht]
\subfigure[]{    \includegraphics[width=0.47\textwidth,clip]{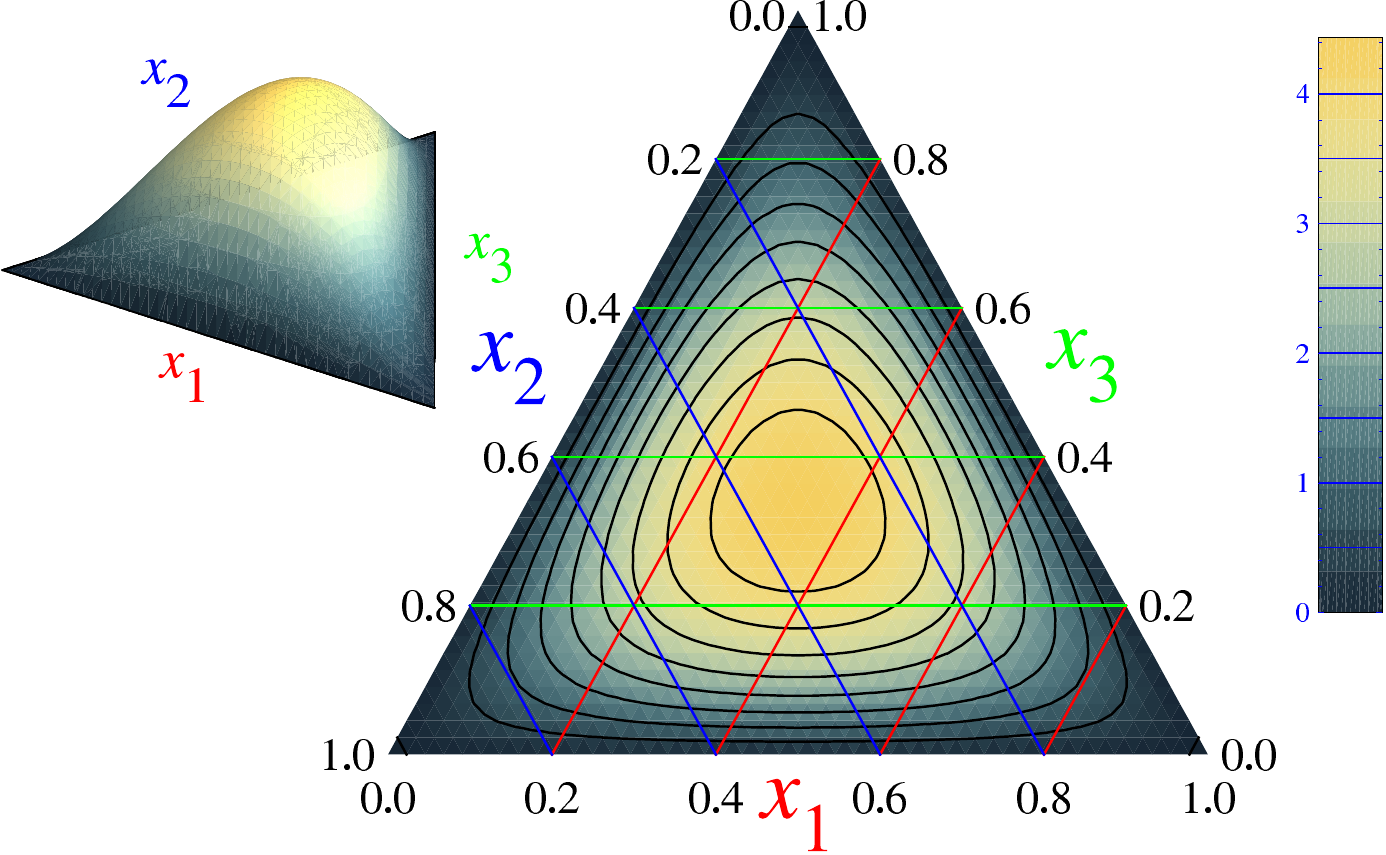}
               \label{fig_ndapics_as}
          }
\subfigure[]{    \includegraphics[width=0.47\textwidth,clip]{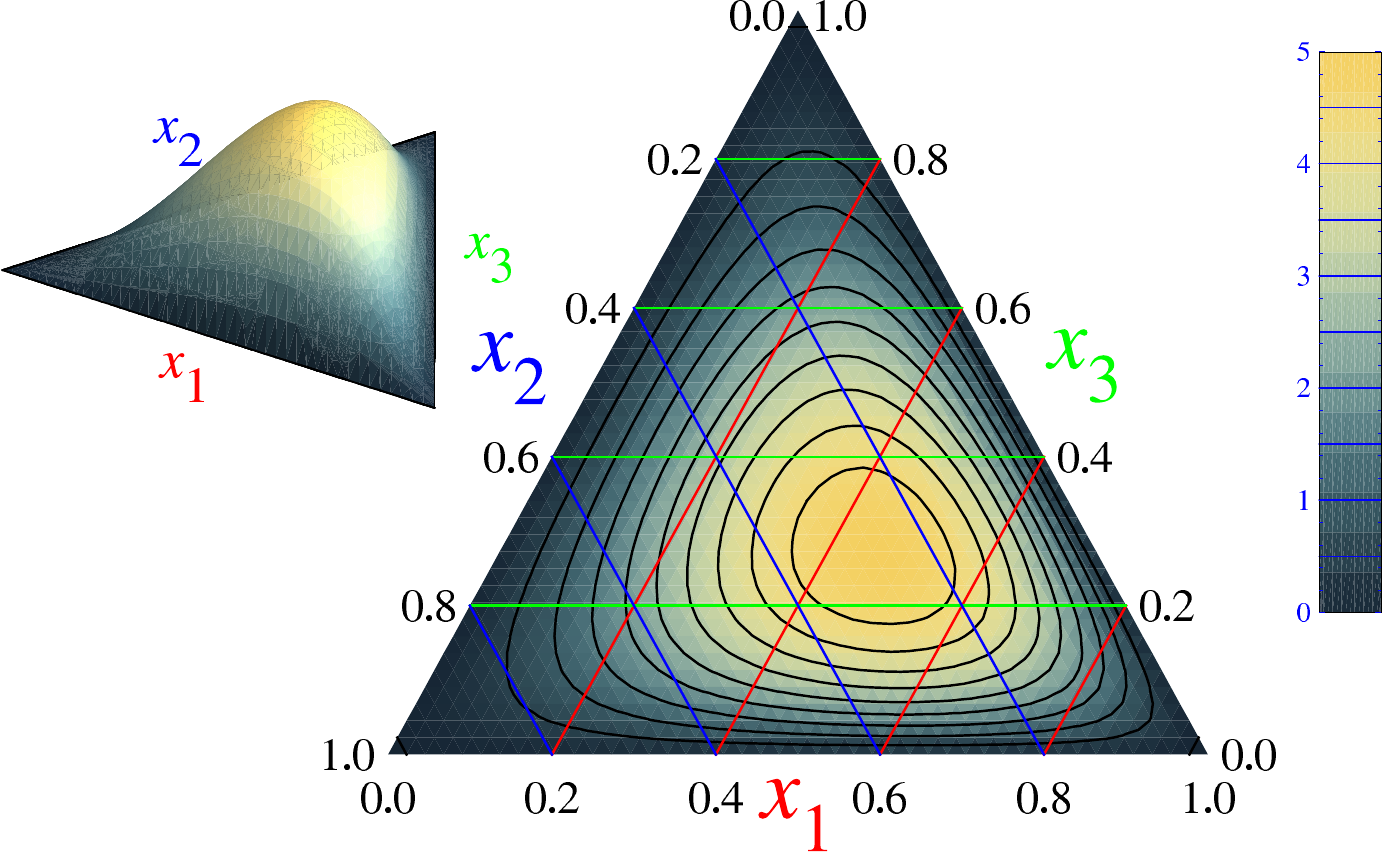}
               \label{fig_ndapics_1}
          }\\
\subfigure[]{    \includegraphics[width=0.47\textwidth,clip]{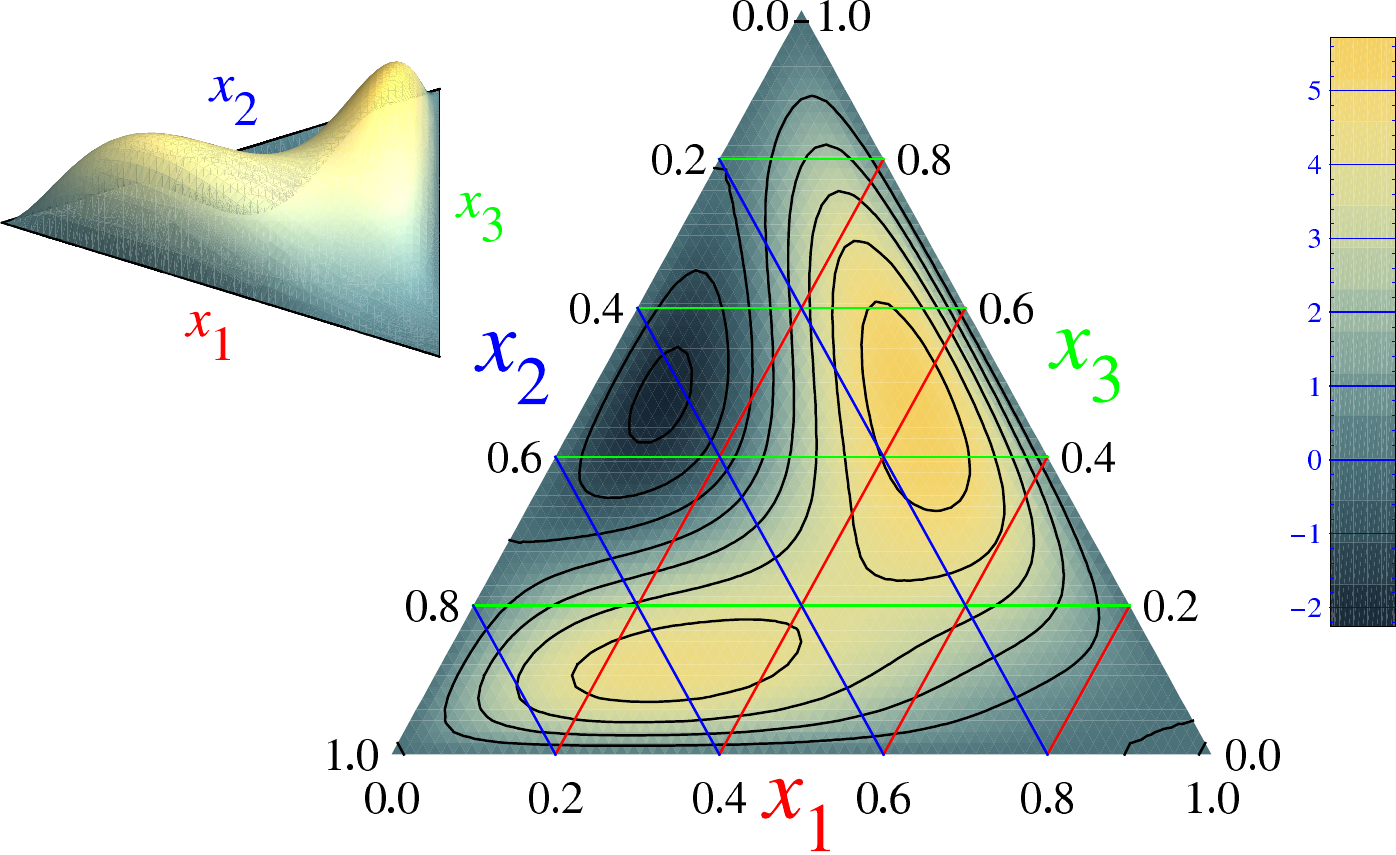}
               \label{fig_ndapics_2}
          }
\subfigure[]{    \includegraphics[width=0.47\textwidth,clip]{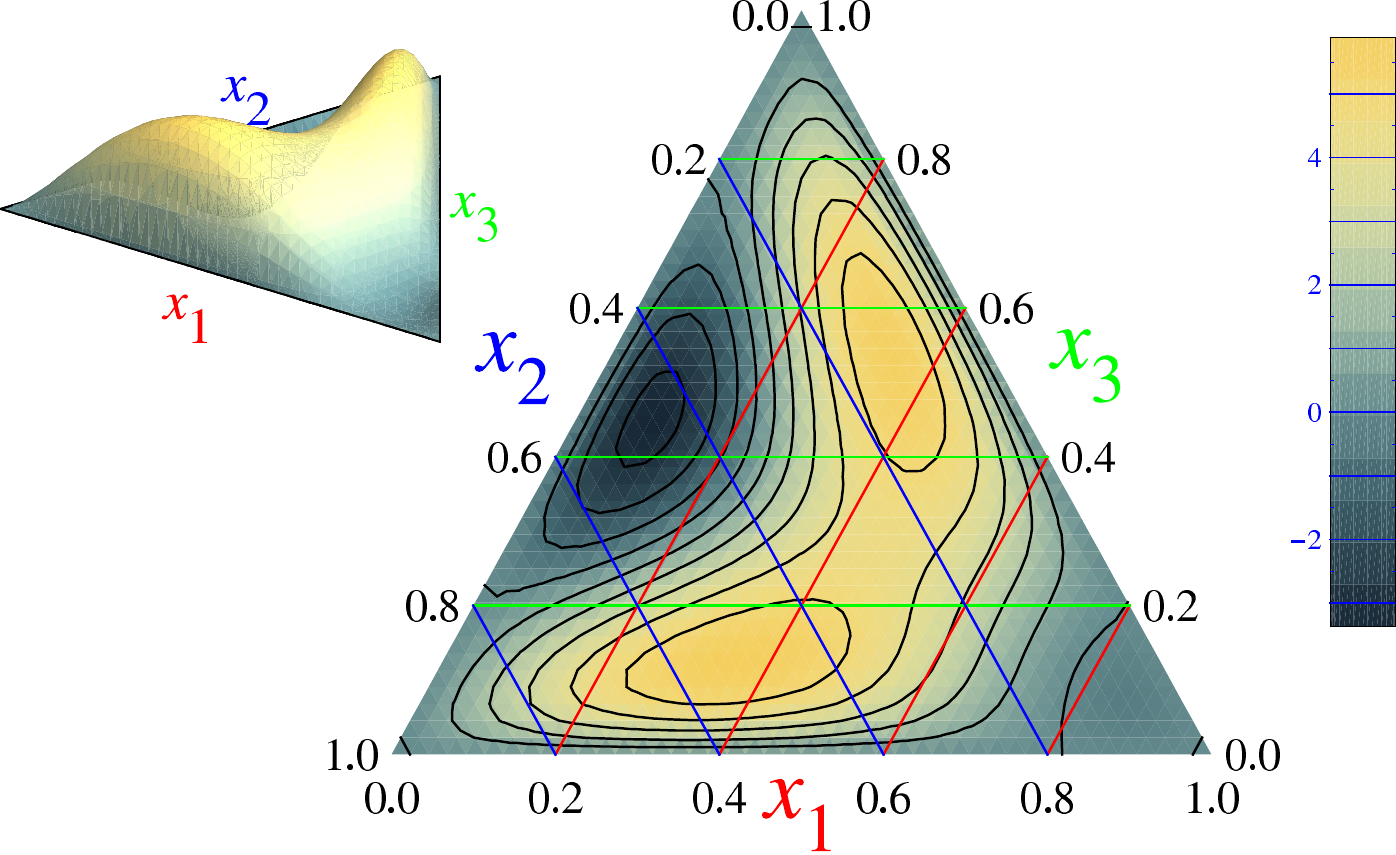}
               \label{fig_ndapics_2a}
          }
 \caption{ \label{fig_ndapics}
 Barycentric contour plot of the leading-twist distribution amplitude
 $\varphi(x_1,x_2,x_3,\mu^2)$ in the
limit of $Q^2=\mu^2\rightarrow \infty$ (a) and at $\mu^2=4\,\mathrm{GeV}^2$
(b-d) using expansion \eqref{eq_daexpansion} as obtained from the $\beta=5.40$
moments presented in Table~\ref{tab_chi_cf}. The asymmetry caused by the first
moments only $(N=1)$ is illustrated in (b), while in (c-d) we took into account
also the second moments $(N=2)$. In (c) we have used set 1 and in (d) set 2 as 
described in the text. The lines of constant $x_1$, $x_2$ and $x_3$ are
parallel to the sides of the triangle labelled by $x_2$, $x_3$ and $x_1$,
respectively.
}
\end{figure}

Our final results for the coefficients 
$c_{nl}$ at the renormalization scale $\mu^2=4\,\mathrm{GeV}^2$
as obtained from the $\beta=5.40$ moments presented in 
Table~\ref{tab_chi_cf}
are collected in Table~\ref{tab:c_nl}.
As the central values for the moments $\phi^{lmn}$ with $l+m+n=2$ 
do not fulfill the constraint
\eqref{eq:sumrule} exactly, the values of $c_{20}$,  $c_{21}$,  $c_{22}$ depend
on the set of moments $\phi^{lmn}$ that are used as an input.
To illustrate this effect, we show two sets of the coefficients
obtained from $\phi^{101}$, $\phi^{200}$, $\phi^{002}$ (set 1) 
and $\phi^{101}$, $\phi^{011}$, $\phi^{110}$ (set 2).
The difference between the two sets is, of course, part of the 
uncertainty of the calculation. We estimate the overall uncertainty
to be about 30\% for  $c_{10}$, $c_{11}$, of order 50\% 
for  $c_{20}$,  $c_{21}$ and a factor of two for $c_{22}$.
 
\begin{table}[ht]
\centering 
\renewcommand{\arraystretch}{1.25}
\begin{tabular}{ c d d }
\hline\hline
                     & \multicolumn{1}{c}{Set 1}  & \multicolumn{1}{c}{Set 2} \\ 
\hline
  $c_{10}$  &  0.326        &  0.326   \\
  $c_{11}$  &  0.940        &  0.940   \\
\hline
  $c_{20}$  &  -0.872      & -0.687   \\
  $c_{21}$  &  -3.130      & -5.210   \\
  $c_{22}$  &  0.405       &  0.036   \\
\hline
\end{tabular}
\caption{ \label{tab:c_nl}
Central values of the coefficients $c_{nl}$ in the expansion (\ref{eq:model})  
at the renormalization scale $\mu^2=4\,\mathrm{GeV}^2$
as obtained from the $\beta=5.40$ moments presented in 
Table~\ref{tab_chi_cf}. 
}
\end{table}

The resulting shape of the nucleon DA is illustrated in 
Fig.~\ref{fig_ndapics}. The asymptotic DA 
corresponding to the leading term in the expansion (\ref{eq:model}) 
is shown in Fig.~\ref{fig_ndapics_as}. It is totally symmetric 
in the three quark momentum fractions. The model obtained by 
adding the terms proportional to $c_{10}$ and $c_{11}$ is presented    
in Fig.~\ref{fig_ndapics_1}. Compared to the asymptotic case, 
the maximum is shifted towards larger values of $x_1$ indicating that the
first quark carries a larger fraction of the proton momentum.
Finally, for the plots in Figs.~\ref{fig_ndapics_2} and 
\ref{fig_ndapics_2a} we add 
contributions of the second order polynomials $(n=2)$, using the 
coefficients $c_{20}$, $c_{21}$, $c_{22}$ from the first and the second 
set in Table~\ref{tab:c_nl}, respectively. 
The difference is in fact not too large and
the effect is the same in both cases: The maximum is smeared out forming
two local maxima and one local minimum. While the model function from set 2 
exhibits an approximate symmetry 
$\varphi(x_1,x_2,x_3)\approx\varphi(x_1,x_3,x_2)$, 
this property is less obvious in the case of set 1. 
However, the general pattern is preserved.

Whether the change in the shape of the DA caused by adding the 
second-order polynomials is of phenomenological significance
can only be investigated in a dedicated study, which goes beyond 
the scope of this work.
Note, however, that in full analogy to 
usual quantum mechanics, the quality of an approximation to the 
wave function has to be measured with respect to the scalar 
product of the appropriate Hilbert space, in our case Eq.~(\ref{measure}), 
and not pointwise in, e.g., the momentum fraction
representation.   

\begin{figure}[t]    
     \includegraphics[width=0.59\textwidth,clip]{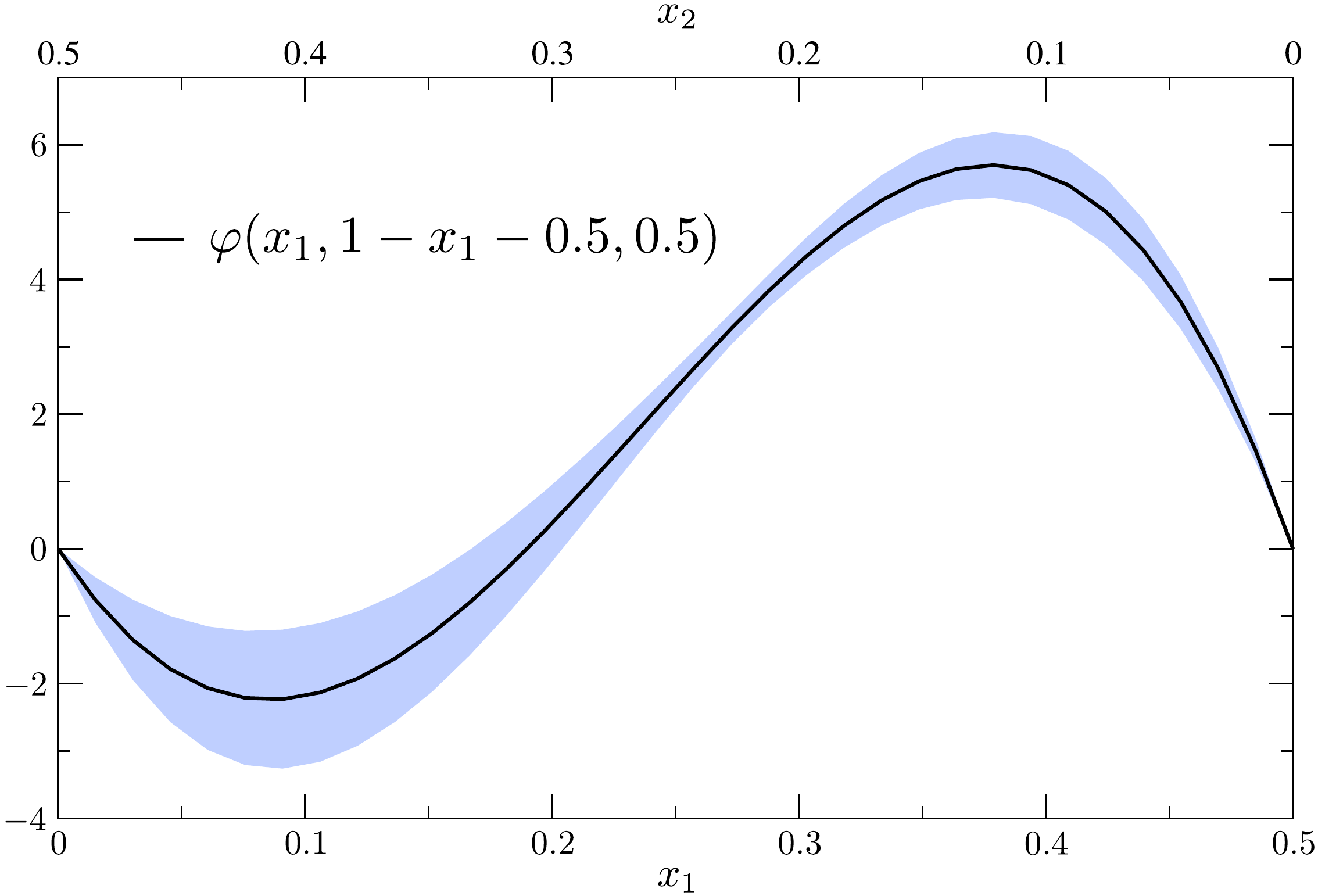}
\caption{  \label{fig_errorprofile} 
The model function $\varphi(x_i)$ for $x_1$ at
$x_3=0.5$ with its statistical uncertainty.
}
\end{figure}

\begin{figure}[t] 
     \includegraphics[width=0.59\textwidth,clip]{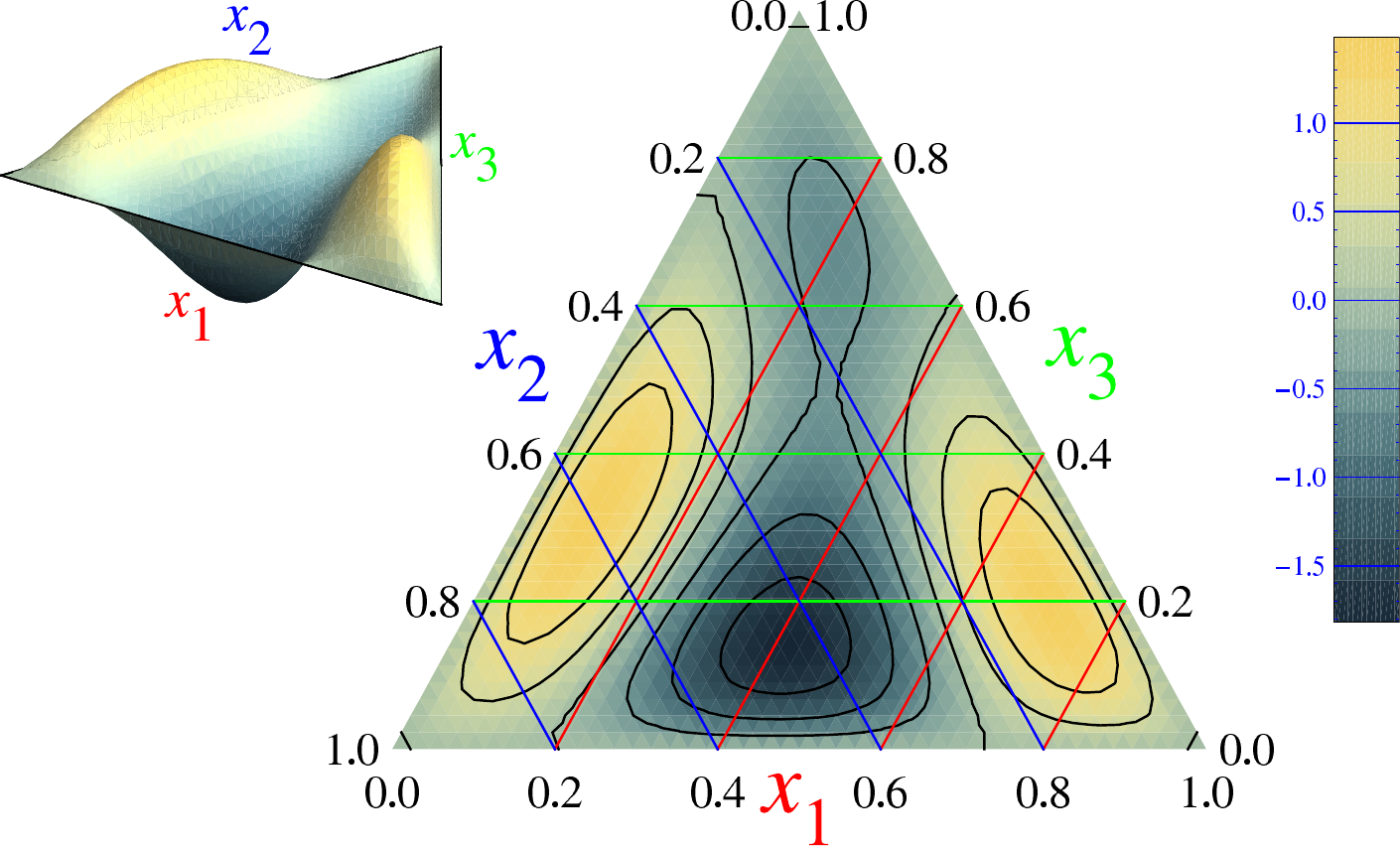}
\caption{ \label{fig_ndadiff}
Systematic uncertainty due to the choice of the independent
subsets of $\phi^{lmn}$ with $l+m+n=2$ (for details see text). The lines of
constant $x_1$, $x_2$ and $x_3$ are parallel to the sides of the triangle
labelled by $x_2$, $x_3$ and $x_1$, respectively.
}
\end{figure}

In Fig.~\ref{fig_errorprofile} we show $\varphi(x_i)$ at $x_3=0.5$ 
as a function of $x_1\, (x_2) $ together with the statistical 
error in order to give an impression of the corresponding uncertainty. 
The effect of choosing different subsets of $\phi^{lmn}$ with 
$l+m+n=2$ is demonstrated in Fig.~\ref{fig_ndadiff} 
where we plot the difference of $\varphi(x_i)$ for set 1 and set 2.

\section{Summary and Conclusions}
\label{sec:summary}

We have evaluated the first few moments of the leading-twist
nucleon DA in lattice QCD. Along with these moments we have determined
the nucleon couplings to local subleading (twist-four) operators.
The required correlators have been computed on gauge field
configurations generated by the QCDSF/DIK collaborations using
two dynamical flavors of clover fermions. The necessary renormalization
matrices have been calculated nonperturbatively, including the 
mixing with operators containing total derivatives.

We have worked with two different gauge couplings corresponding
to $\beta = 5.29$ and $\beta = 5.40$. For the lattice sizes and 
quark masses see Table~\ref{tab_latset}. As our final numbers
we take the data from our finer lattice ($\beta=5.40$).  
The results for the moments of the leading-twist DA
at two different renormalization scales are presented in 
Table~\ref{tab:philmn}. The corresponding coefficients in the 
expansion of the DA in a basis of orthogonal polynomials are 
given in Table~\ref{tab:c_nl}. Truncating this expansion at the 
second order, we obtain a model of the DA which is plotted in 
Fig.~\ref{fig_ndapics}. Our error estimates include statistic and 
known sources of systematic uncertainties, but still have to be 
considered with some caution. The largest uncertainty is caused by
the chiral extrapolation. We expect that it will be reduced in
the relatively near future when simulations with smaller pion masses 
on larger lattices become available.
 
Our value for the coupling $f_N$, which determines the normalization of 
the leading-twist nucleon DA, appears to be approximately 
40\% below the corresponding QCD sum rule 
estimates~\cite{Chernyak:1984bm,King:1986wi,Chernyak:1987nu}. 
If confirmed, this result would deal yet another blow at the hopes 
to calculate the nucleon magnetic form factor
at realistic momentum transfers within perturbative QCD. At the same time,
the twist-four couplings $\lambda_1$ and $\lambda_2$, which are related to 
the normalization of subleading twist-four DAs, turn out to be in 
agreement with other estimates. These constants are relevant, e.g., 
for the description of form factors involving a helicity 
flip within perturbative QCD~\cite{Belitsky:2002kj} 
and also for soft (end-point) corrections to the form factors in the 
light-cone sum rule approach \cite{Braun:2001tj,Braun:2006hz}.   
The same constants enter the effective baryon chiral Largangian 
and can be used to estimate the proton life time within GUT models.

The results we have obtained for the first moments of the nucleon 
DA are consistent with the conventional picture that the valence $u$-quark 
with helicity parallel to that of the proton carries 
the largest fraction of its momentum, but the effect seems to 
be less pronounced compared to the corresponding QCD sum rule 
calculations~\cite{Chernyak:1984bm,King:1986wi,Chernyak:1987nu}. 
Our numbers, however, are compatible with those extracted from 
the fits to the electromagnetic proton form factors within the light-cone 
sum rule approach \cite{Braun:2006hz}.

Our calculation of the second moments of the DA indicates the presence of 
considerable second-order contributions in the expansion
in terms of orthogonal polynomials. Qualitatively, these
contributions smear out the maximum forming
two local maxima and one local minimum (see Figs.~\ref{fig_ndapics_2} 
and \ref{fig_ndapics_2a}). The investigation of the phenomenological 
consequences of these and other features of our model DA, 
such as the approximate symmetry 
$\varphi(x_1,x_2,x_3)\approx\varphi(x_1,x_3,x_2)$,
requires a dedicated study, which goes beyond the scope of the 
present work and will be presented elsewhere.

\begin{acknowledgments}
 We are grateful to A.~Lenz, J.~Bloch and A.~Manashov for helpful
discussions. The numerical calculations have been performed on the Hitachi
SR8000 at LRZ (Munich), apeNEXT and APEmille at NIC/DESY (Zeuthen) and
BlueGene/Ls at NIC/JSC (J\"ulich), EPCC (Edinburgh) and KEK (by the Kanazawa
group as part of the DIK research program) as well as  QCDOC (Regensburg)
 using the Chroma software library \cite{Edwards:2004sx, bagel:2005}. 
This work was supported by DFG
(Forschergruppe Gitter-Hadronen-Ph\"anomenologie and SFB/TR55 Hadron Physics 
from Lattice QCD), by EU I3HP (contract No. RII3-CT-2004-506078)  and by BMBF.
\end{acknowledgments}

\clearpage
\appendix
\squeezetable
\section{Dirac matrices in Weyl representation \label{app:weyl}}
We have used the following representation of the Euclidean Dirac matrices:
   \begin{align}
      \gamma_1&=
      \left(
         \begin{array}{llll}
          0 & 0 & 0 & i \\
          0 & 0 & i & 0 \\
          0 & -i & 0 & 0 \\
          -i & 0 & 0 & 0
         \end{array}
      \right), &
      \gamma_2&=
      \left(
         \begin{array}{llll}
          0 & 0 & 0 & 1 \\
          0 & 0 & -1 & 0 \\
          0 & -1 & 0 & 0 \\
          1 & 0 & 0 & 0
         \end{array}
      \right), &
      \gamma_3 & =\left(
         \begin{array}{llll}
          0 & 0 & i & 0 \\
          0 & 0 & 0 & -i \\
          -i & 0 & 0 & 0 \\
          0 & i & 0 & 0
         \end{array}
      \right), &
      \gamma_4&=
      \left(
         \begin{array}{llll}
          0 & 0 & 1 & 0 \\
          0 & 0 & 0 & 1 \\
          1 & 0 & 0 & 0 \\
          0 & 1 & 0 & 0
         \end{array}
       \right) 
   \end{align} 
with
\begin{equation}
      \gamma_5 = \gamma_1 \gamma_2 \gamma_3 \gamma_4 =
      \left(
         \begin{array}{llll}
          -1 & 0 & 0 & 0 \\
          0  & -1& 0 & 0 \\
          0  & 0 & 1 & 0 \\
          0  & 0 & 0 & 1
         \end{array}
       \right), \qquad 
	\sigma_{\mu\nu}
=\frac{i}{2}\left(\gamma_\mu\gamma_\nu-\gamma_\nu\gamma_\mu\right) \,.
\end{equation}
The charge conjugation matrix has been chosen as
\begin{equation}
  C=\gamma_2\gamma_4\,.
\end{equation}

\section{Operator relations for leading-twist distribution 
amplitudes \label{app:mom}}

In the following we give the relations between the operators whose
matrix elements define moments of the leading-twist DA of spin-1/2
baryons (DA operators) and the irreducible operators that appear in the
general group-theoretical classification in \cite{Kaltenbrunner:2008pb}.  
The relations are written for general quark flavors $f$, $g$, $h$; 
the proton case is obtained by the replacement $f,g\to u$, $h\to d$ and the 
appropriate symmetrization to single out the contribution of isospin 1/2.  

The total symmetrization in space-time indices denoted by the curly brackets,
e.g., 
$$\mathcal V^{\{23\}}=\frac{1}{2!}(\mathcal V^{23}+\mathcal V^{32})$$ 
reflects the leading-twist projection. 
For example, the moment  $V^{001}$ is calculated from
\begin{equation}
  \frac{1}{2!} \epsilon^{a b c} \;
	\left(
                                      \left[f(0)\right]^a_\alpha 
                                          \;     (C\gamma_2)_{\alpha\beta}      
\;
                                       \left[ g(0) \right]^b_\beta \;
                                       \left[ i D_3 (\gamma_5 h(0))
\right]^c_\tau
	+
                                      \left[f(0)\right]^a_\alpha 
                                          \;     (C\gamma_3)_{\alpha\beta}      
\;
                                       \left[ g(0) \right]^b_\beta \;
                                       \left[ i D_2 (\gamma_5 h(0) )
\right]^c_\tau
	\right).
\end{equation}
In the notation used below, it is not indicated explicitly on which quark the
derivatives act in the operators on the right-hand side. However, it is always
implied that the positions of the derivatives are the same on both sides of the
equations.

\subsection*{0th moment \label{app_mom0}}

     \begin{align}
      \left( \mathcal B^{000}_{9,6},  -\mathcal B^{000}_{9,1}, -\mathcal
B^{000}_{9,12}, \mathcal B^{000}_{9,7} \right)=&
      \frac{1}{4}
      \left(
         \gamma_3\gamma_4\left[  \gamma_2 \mathcal{T}^{1} 
                                +\gamma_1 \mathcal{T}^{2} 
                         \right]
      \right)\\
      \left( \mathcal B^{000}_{9,4},  -\mathcal B^{000}_{9,3}, -\mathcal
B^{000}_{9,10}, \mathcal B^{000}_{9,9} \right)=&
      \frac{1}{4}
      \left(
         \gamma_1\gamma_2\left[  \gamma_4 \mathcal{T}^{3} 
                                +\gamma_3 \mathcal{T}^{4} 
                         \right]
      \right)\\
      \left( \mathcal B^{000}_{9,2},  -\mathcal B^{000}_{9,5}, -\mathcal
B^{000}_{9,8}, \mathcal B^{000}_{9,11} \right)=&
	\frac{1}{4\sqrt{2}}
      \left(
         \gamma_1\gamma_2\left[  \gamma_4 \mathcal{T}^{3} 
                                -\gamma_3 \mathcal{T}^{4} 
                         \right]
        +\gamma_3\gamma_4\left[  \gamma_1 \mathcal{T}^{2} 
                                -\gamma_2 \mathcal{T}^{1} 
                         \right]
      \right)
   \end{align}
The $\mathcal B^{000}_{7,i}$  ( $\mathcal B^{000}_{8,i}$) operators from the
symmetry class $-++$ $(+-+)$ are obtained from the above operators by replacing
$\mathcal T$ on the right hand side by $\mathcal V + \mathcal A$ ($\mathcal V -
\mathcal A$).

\subsection*{1st moments  \label{app_mom1}}

   \begin{align}
      \left( \mathcal B^{lmn}_{7,1},  -\mathcal B^{lmn}_{7,2}, \mathcal B^{lmn}_{7,7}, -\mathcal B^{lmn}_{7,8} \right)=&
       \frac{1}{4\sqrt{2}}
      \left(
         2\gamma_4\gamma_3  \mathcal{T}^{\{12\}}
         +\gamma_4\gamma_2  \mathcal{T}^{\{13\}}
         +\gamma_2\gamma_3  \mathcal{T}^{\{14\}}
         +\gamma_4\gamma_1  \mathcal{T}^{\{23\}}
         +\gamma_1\gamma_3  \mathcal{T}^{\{24\}}
      \right)\\
      \left( \mathcal B^{lmn}_{7,3},  -\mathcal B^{lmn}_{7,4}, \mathcal B^{lmn}_{7,9}, -\mathcal B^{lmn}_{7,10} \right)=&
       \frac{1}{4\sqrt{2}}
      \left(
         2\gamma_1\gamma_2  \mathcal{T}^{\{34\}}
         +\gamma_4\gamma_2  \mathcal{T}^{\{13\}}
         +\gamma_3\gamma_2  \mathcal{T}^{\{14\}}
         +\gamma_1\gamma_4  \mathcal{T}^{\{23\}}
         +\gamma_1\gamma_3  \mathcal{T}^{\{24\}}
      \right)\\
      \left(  \mathcal B^{lmn}_{7,6},  \mathcal B^{lmn}_{7,5}, \mathcal B^{lmn}_{7,12}, \mathcal B^{lmn}_{7,11} \right)=&
       \frac{1}{4}
      \left(
          \gamma_2\gamma_4  \mathcal{T}^{\{13\}}
         +\gamma_2\gamma_3  \mathcal{T}^{\{14\}}
         +\gamma_1\gamma_4  \mathcal{T}^{\{23\}}
         +\gamma_1\gamma_3  \mathcal{T}^{\{24\}}
      \right)
   \end{align}
The $\mathcal B^{lmn}_{5,i}$  ( $\mathcal B^{lmn}_{6,i}$) operators from the symmetry class $D-++$ $(D+-+)$ are obtained from the above operators by replacing $\mathcal T$ on the right hand side by $\mathcal V + \mathcal A$ ($\mathcal V - \mathcal A$).

\subsection*{2nd moments \label{app_mom2}}

   \begin{align}
      \left( -\mathcal B^{lmn}_{6,4},  -\mathcal B^{lmn}_{6,3}, \mathcal B^{lmn}_{6,2}, \mathcal B^{lmn}_{6,1} \right)=&
      \frac{\sqrt{3}}{4}
      \left(
          \gamma_4  \mathcal{T}^{\{123\}}
         +\gamma_3  \mathcal{T}^{\{124\}}
         +\gamma_2  \mathcal{T}^{\{134\}}
         +\gamma_1  \mathcal{T}^{\{234\}}
      \right)
   \end{align}
The $\mathcal B^{lmn}_{4,i}$  ( $\mathcal B^{lmn}_{5,i}$) operators from the symmetry class $DD-++$ $(DD+-+)$ are obtained from the above operators by replacing $\mathcal T$ on the right hand side by $\mathcal V + \mathcal A$ ($\mathcal V - \mathcal A$).

\section{Raw lattice results \label{app:latres}}

In this appendix we collect the results of the linear (in $m_\pi^2$) 
extrapolation of our bare lattice data. The errors given are purely statistical.

\begin{table}[ht]
\centering
\footnotesize
\renewcommand{\arraystretch}{1.25}
   \begin{tabular*}{0.97\textwidth}{@{\extracolsep{\fill} } c D{.}{.}{9} D{.}{.}{4}  D{.}{.}{9} D{.}{.}{4}  D{.}{.}{9} D{.}{.}{4} }
\hline\hline
       &   \multicolumn{4}{c}{ $\beta=5.29$ }      &   \multicolumn{2}{c}{ $\beta=5.40$ } \\
\hline
                 &   \multicolumn{2}{c}{ all }      &  \multicolumn{2}{c}{ 24 }  &   \multicolumn{2}{c}{ all } \\ 
\hline
                 & \multicolumn{1}{c}{\#}& \multicolumn{1}{c}{\dev}&\multicolumn{1}{c}{\#}&\multicolumn{1}{c}{\dev} &\multicolumn{1}{c}{\#}&\multicolumn{1}{c}{\dev}  \\
\hline
$ f_N/m_N^2 \cdot 10^{3}$                    & 4.088(77)   & 6.563 &  4.53(14)    & 0.555   & 4.287(74)   & 0.658 \\
$-\lambda_1/m_N \cdot10^{3}[\mathrm{GeV}]$   & 60.80(106)  & 19.31 &  69.28(176)  & 6.209   & 59.40(95)   & 1.060 \\
$-\lambda_1 \cdot 10^{3}[\mathrm{GeV}^2]$    & 77.33(149)  & 18.46 &  82.24(209)  & 3.484   & 72.86(135)  & 1.901 \\
$ \lambda_2/m_N \cdot 10^{3} [\mathrm{GeV}]$ & 129.76(214) & 19.98 & 141.53(360)  & 4.928   &119.16(191)  & 1.498 \\
$ \lambda_2 \cdot 10^{3} [\mathrm{GeV}^2]$   & 158.00(315) & 18.31 & 168.30(428)  & 2.388   &146.48(270)  & 2.716 \\
\hline
  $\phi^{100}$                               & 0.2987(49) & 1.125 &    0.315(10)  & 0.033   & 0.2939(59)  & 1.384 \\
  $\phi^{010}$                               & 0.2746(48) & 0.768 &    0.263(11)  & 0.765   & 0.2719(62)  & 0.335 \\
  $\phi^{001}$                               & 0.2840(48) & 1.566 &    0.271(11)  & 2.555   & 0.2740(60)  & 0.972 \\
\hline
  $\phi^{011}$                               & 0.0647(37) & 0.276 &    0.0633(87) & 0.711   & 0.0646(44)  & 1.831 \\
  $\phi^{101}$                               & 0.0606(39) & 0.821 &    0.067(12)  & 0.744   & 0.0688(55)  & 1.057 \\
  $\phi^{110}$                               & 0.0651(32) & 0.712 &    0.0592(79) & 0.445   & 0.0707(39)  & 0.610 \\
  $\phi^{200}$                               & 0.1149(54) & 2.367 &    0.146(14)  & 0.597   & 0.1126(68)  & 5.534 \\
  $\phi^{020}$                               & 0.0922(50) & 0.717 &    0.096(12)  & 1.908   & 0.0949(61)  & 0.288 \\
  $\phi^{002}$                               & 0.1067(54) & 0.944 &    0.108(13)  & 2.729   & 0.1060(64)  & 0.114 \\
\hline
\end{tabular*}
\caption{ 
Linear extrapolations of FC (unconstrained) results to the physical point using
all available lattice ensembles (all) and  $24^3\times 48$ lattices only (24) 
for $\beta=5.29$. The \dev\ refers to the linear chiral extrapolation.
}
\end{table}

\begin{table}[ht]
\centering
\renewcommand{\arraystretch}{1.19}
   \begin{tabular*}{0.97\textwidth}{@{\extracolsep{\fill} } c D{.}{.}{9} D{.}{.}{4}   D{.}{.}{9} D{.}{.}{4}  D{.}{.}{9} D{.}{.}{4} }
\hline\hline
       &   \multicolumn{4}{c}{ $\beta=5.29$ }      &   \multicolumn{2}{c}{ $\beta=5.40$ } \\
\hline
                 &   \multicolumn{2}{c}{ all }      &  \multicolumn{2}{c}{ 24 }  &   \multicolumn{2}{c}{ all } \\ 
\hline
                 & \multicolumn{1}{c}{\#}& \multicolumn{1}{c}{\dev}&\multicolumn{1}{c}{\#}&\multicolumn{1}{c}{\dev}   \\
\hline
$ f_N/m_N^2 \cdot 10^{3}$          & 4.396(99)   & 2.417  & 4.67(19)  & 1.208   & 4.517(96)   & 0.342  \\
\hline
$V^{100}=V^{010}$                  & 0.308(13)   & 0.416 &  0.298(35) & 0.027   & 0.298(19)   & 0.966  \\
$A^{100}=-A^{010}$                 & 0.0133(40)  & 2.495 &  0.046(13) & 0.038   & 0.0196(64)  & 0.960  \\
$T^{100}=T^{010}$                  & 0.307(12)   & 0.425 &  0.297(25) & 0.263   & 0.300(16)   & 0.483  \\

$\varphi^{100}$                    & 0.324(16)   & 0.352 &  0.360(49) & 0.001   & 0.323(24)   & 0.777  \\
$\varphi^{010}=\phi^{010}=T^{001}$ & 0.286(12)   & 1.636 &  0.248(26) & 0.550   & 0.276(17)   & 0.446  \\
$\varphi^{001}=V^{001}$            & 0.289(15)   & 1.892 &  0.229(37) & 1.532   & 0.280(21)   & 0.399  \\

$\phi^{100}-\phi^{010}$            & 0.0194(49)  & 2.230 &  0.054(15) & 0.056   & 0.0258(77)  & 0.928  \\
$\phi^{100}-\phi^{001}$            & 0.0076(39)  & 2.017 &  0.036(14) & 1.011   & 0.0129(66)  & 1.291  \\
$\phi^{001}-\phi^{010}$            & 0.0114(41)  & 0.679 &  0.016(13) & 1.719   & 0.0144(66)  & 2.118  \\
\hline
$V^{011}=V^{101}$                  & 0.0698(56)  & 0.197 &  0.072(17) & 0.228   & 0.0676(69)  & 0.260  \\
$A^{011}=-A^{101}$                 & -0.0006(49) & 0.038 &  0.000(15) & 0.004   & 0.0022(60)  & 1.063  \\
$T^{011}=T^{101}$                  & 0.0689(44)  & 0.395 &  0.068(12) & 0.035   & 0.0707(54)  & 0.580  \\

$\varphi^{011}$                    & 0.0709(85)  & 0.068 &  0.076(27) & 0.061   & 0.064(11)   & 0.533  \\
$\varphi^{101}=\phi^{101}=T^{110}$ & 0.0699(62)  & 0.428 &  0.071(18) & 0.135   & 0.0673(67)  & 0.504  \\
$\varphi^{110}=V^{110}$            & 0.0637(79)  & 0.149 &  0.064(24) & 0.101   & 0.077(10)   & 0.049  \\

$\phi^{101}-\phi^{011}$            & 0.0012(62)  & 0.068 &  0.006(19) & 0.023   & 0.0005(73)  & 1.711  \\
$\phi^{011}-\phi^{110}$            & 0.0025(45)  & 0.048 &  0.004(15) & 0.096   & -0.0042(62) & 0.246  \\
$\phi^{101}-\phi^{110}$            & -0.0001(47) & 0.155 &  0.005(17) & 0.383   & -0.0036(62) & 0.627  \\
\hline
$V^{200}=V^{020}$                  & 0.1059(78)  & 0.557 &  0.129(22) & 0.015   & 0.115(10)   & 2.034  \\
$A^{020}=-A^{200}$                 & 0.0132(59)  & 0.698 &  0.036(18) & 0.131   & 0.0195(81)  & 1.812  \\
$T^{200}=T^{020}$                  & 0.1108(79)  & 0.576 &  0.119(19) & 1.336   & 0.1203(89)  & 1.450  \\

$\varphi^{200}$                    & 0.117(12)   & 0.739 &  0.165(37) & 0.006   & 0.134(16)   & 2.305  \\
$\varphi^{020}=\phi^{020}=T^{002}$ & 0.0913(73)  & 0.261 &  0.097(19) & 0.590   & 0.0963(93)  & 0.646  \\
$\varphi^{002}=V^{002}$            & 0.096(12)   & 0.724 &  0.066(35) & 1.320   & 0.106(15)   & 0.279  \\

$\phi^{200}-\phi^{020}$            & 0.0206(68)  & 0.406 &  0.039(21) & 0.001   & 0.0300(97)  & 1.864  \\
$\phi^{200}-\phi^{002}$            & 0.0060(61)  & 0.847 &  0.032(20) & 0.601   & 0.0092(83)  & 1.380  \\
$\phi^{002}-\phi^{020}$            & 0.0114(55)  & 0.291 &  0.005(19) & 0.757   & 0.0215(80)  & 0.438  \\
\hline
\end{tabular*}
\caption{
Linear extrapolations of PC (unconstrained) results to the physical point using
all available lattice ensembles (all) and  $24^3\times 48$ lattices only (24) 
for $\beta=5.29$. The \dev\ refers to the linear chiral extrapolation.
}
\end{table}

\begin{table}[ht]
\renewcommand{\arraystretch}{1.25}
\centering
   \begin{tabular*}{0.97\textwidth}{@{\extracolsep{\fill} } c D{.}{.}{18} D{.}{.}{18}   D{.}{.}{18} D{.}{.}{18}}
\hline\hline
        &\multicolumn{2}{c}{$\beta=5.29$}& \multicolumn{2}{c}{$\beta=5.40$}\\
\hline 
                                              &\multicolumn{1}{c}{\#}& \multicolumn{1}{c}{\dev} &\multicolumn{1}{c}{\#}& \multicolumn{1}{c}{\dev}\\
\hline
$ f_N/m_N^2 \cdot 10^{3}$                     & 4.215(85)        & 1.878  & 4.395(85)        & 0.267\\
$-\lambda_1/m_N \cdot10^{3}[\mathrm{GeV}]$    & 51.10(117)       & 10.57  & 60.35(108)       & 0.184\\
$ \lambda_2/m_N \cdot10^{3}[\mathrm{GeV}]$    & 125.75(25)       & 10.54  & 120.80(216)      & 0.403\\
\hline
  $\phi^{100}$                                & 0.3286(12)       & 7.559  & 0.3358(11)       & 6.115\\
  $\phi^{010}=\varphi^{010}$                  & 0.2943(9)        & 8.530  & 0.2891(9)        & 6.960\\
  ${\phi^{001}}(\star)$                       & 0.3164(9)        & 1.112  & 0.3155(9)        & 1.312\\
\hline
  $\phi^{100}-\phi^{010}$                     & 0.0350(20)       & 9.960  & 0.0468(19)       & 7.732\\
  $\phi^{100}-\phi^{001}$                     & 0.0126(19)       & 3.996  & 0.0206(18)       & 3.300\\
  $\phi^{001}-\phi^{010}$                     & 0.0225(14)       & 3.315  & 0.0263(14)       & 2.526\\
\hline
  $\phi^{011}$                                & 0.1113(26)       & 3.593  & 0.0932(19)       & 1.544\\
  $\phi^{101}$                                & 0.1148(26)       & 0.370  & 0.1124(18)       & 0.287\\
  ${\phi^{110}}(\star)$                       & 0.1085(22)       & 1.716  & 0.1034(16)       & 0.135\\
  $\phi^{200}$                                & 0.1820(44)       & 4.176  & 0.1924(30)       & 0.338\\
  ${\phi^{020}=\varphi^{020}}(\star)$         & 0.1489(35)       & 0.363  & 0.1539(28)       & 0.265\\
  $\phi^{002}$                                & 0.1728(42)       & 1.677  & 0.1801(36)       & 0.856\\
\hline
  $\phi^{101}-\phi^{011}$                     & 0.0042(39)       & 2.489  & 0.0200(27)       & 0.900\\
  $\phi^{110}-\phi^{011}$                     & 0.0042(34)       & 0.636  & 0.0100(25)       & 0.775\\
  $\phi^{101}-\phi^{110}$                     & 0.0053(29)       & 1.159  & 0.0094(20)       & 0.257\\
  $\phi^{200}-\phi^{020}$                     & 0.0367(48)       & 1.515  & 0.0364(35)       & 0.514\\
  $\phi^{200}-\phi^{002}$                     & 0.0076(59)       & 1.763  & 0.0115(39)       & 0.810\\
  $\phi^{002}-\phi^{020}$                     & 0.0230(39)       & 1.010  & 0.0255(24)       & 0.597\\
\hline
\end{tabular*}
\caption{ \label{tab_chi_cf_b529}
Linear extrapolations of $\phi^{lmn}$ and asymmetries to the physical point as 
obtained from the constrained analysis using all available ensembles. The \dev\ refers to the linear chiral extrapolation. The values denoted by the $\star$ were used to determine the absolute normalization of the associated asymmetries. 
}
\end{table}

\cleardoublepage

\end{document}